\documentclass[aps,pra,onecolumn,superscriptaddress,notitlepage,nofootinbib,longbibliography, colorlinks=true]{revtex4-2}
\usepackage[utf8]{inputenc}
\usepackage[T1]{fontenc}
\usepackage{array}[=2016-10-06]
\usepackage{dcolumn}
\usepackage{amsmath,amssymb,amsfonts,graphicx,float,mathtools,mathptmx}
\usepackage[colorlinks, linkcolor=blue, citecolor=blue,  breaklinks=true]{hyperref}
\usepackage{amssymb,amsmath,graphicx,bm,float,xcolor,booktabs,siunitx}  
\usepackage[titletoc,title]{appendix}
\usepackage[shortlabels]{enumitem}
\usepackage[capitalize,noabbrev]{cleveref}
\begin{document}
	\title{Multipartite quantum entanglement in $\mathcal{PT}$-symmetric molecular optomechanics: Nonreciprocal enhancement and thermal resilience to \SI{500}{\kelvin}}
	
	\author{E. Kongkui Berinyuy}
	\email{emale.kongkui@facsciences-uy1.cm}
	\affiliation{Department of Physics, Faculty of Science, University of Yaounde I, P.O.Box 812, Yaounde, Cameroon}

	\author{C. Tchodimou}
	\affiliation{Department of Physics, Faculty of Science, University of Yaounde I, P.O.Box 812, Yaounde, Cameroon}

	\author{P. Djorwé}
	\affiliation{Department of Physics, Faculty of Science,
		University of Ngaoundere, P.O.Box 454, Ngaoundere, Cameroon}
	\affiliation{Stellenbosch Institute for Advanced Study (STIAS), Wallenberg Research Centre at Stellenbosch University, Stellenbosch 7600, South Africa}

	\author{Jia-Xin Peng}
		\affiliation{School of Physics and Technology, Nantong University, Nantong, 226019, People’s Republic of China}

		\author{S. K. Singh}
	\affiliation{ Department of Physics, Akal University, Talwandi Sabo, Bathinda 151302, India}

	\author{S. G. Nana Engo}
	\email{serge.nana-engo@facsciences-uy1.cm}
	\affiliation{Department of Physics, Faculty of Science, University of Yaounde I, P.O.Box 812, Yaounde, Cameroon}

\begin{abstract}
We present a theoretical framework for a $\mathcal{PT}$-symmetric double-cavity molecular optomechanical system demonstrating nonreciprocal enhancement of multipartite quantum entanglement at elevated temperatures. All bipartite entanglement channels ($E_{ac}$, $E_{aB_1}$, $E_{cB_2}$, $E_{B_1B_2}$) simultaneously maximize at optimal nonreciprocal asymmetry $J_1/J_2 \approx 5$, with entanglement persisting to $T \sim \SIrange{400}{500}{\kelvin}$ (material-limited ceiling) two orders of magnitude beyond conventional optomechanical systems. This thermal resilience and balanced enhancement across all channels arise from synergistic combination of ultra-high-frequency molecular vibrations ($\omega_m/2\pi = \SI{30}{\tera\hertz}$), collective $\sqrt{N}$ coupling enhancement with $N=\num{e6}$ molecules, and directional nonreciprocal coupling shielding entanglement-generating interactions from backaction noise. Unlike optical parametric amplifier schemes where vibration-vibration enhancement suppresses optical-vibration correlations, our $\mathcal{PT}$-symmetric architecture circumvents this fundamental trade-off, validated through rigorous stability analysis via Routh-Hurwitz criterion.
\end{abstract}

	\maketitle

\section{Introduction} \label{sec:Intro}

In recent years, efforts have been devoted to the generation, control, and protection of quantum entanglement against the effects of environmental noise and the resulting decoherence, as quantum entanglement is a key resource for emerging quantum technologies \cite{Barzanjeh2021, Blais2020, Kotler2021, Xia2023, Brady2023, Chen2025, Ye2025, Araya2023, rogers2014hybrid, Rips2013}. Realizing the full potential of quantum computing, secure communication, and ultra-precise sensing therefore hinges on developing physical platforms where entanglement can be manipulated with precision and resilience \cite{Emale2025, Lai2022, Lai.2022, Berinyuy2025b}.

Molecular cavity optomechanics (McOM) has emerged as a promising platform for quantum state engineering \cite{Roelli2024, Huang2024, Huang2025a}. McOM systems offer three key advantages: (i) ultra-strong light-matter coupling with single-photon optomechanical coupling $g_m/2\pi \sim \SIrange{10}{100}{\mega\hertz}$ reaching collective strengths $G/2\pi \sim \SI{50}{\giga\hertz}$ for $N \sim \num{e6}$ molecules \cite{Roelli2016, Chikkaraddy2016, Xiang2024}, (ii) access to ultra-high-frequency vibrational modes ($\omega_m/2\pi \sim \SIrange{24}{48}{\tera\hertz}$), and (iii) collective $\sqrt{N}$ enhancement of coupling strength \cite{Huang2024, Berinyuy2025a}. Recent demonstrations of coherent Raman scattering in molecular optomechanical cavities \cite{Huang2025a, Zou2024} and theoretical studies of collective quantum entanglement \cite{Huang2024} validate these parameter ranges and enhancement mechanisms. Experimental implementations by Djorwe et al. \cite{Djorwe2022, Djor2024} and theoretical foundations established by Berinyuy et al. \cite{Berinyuy2025, Berinyuy2025a, Berinyuy2025b} support the feasibility of these platforms.

These features provide McOM systems with intrinsic thermal resilience. Ultra-high vibrational frequencies yield low thermal phonon occupation even at elevated temperatures, enabling entanglement to persist theoretically to $T \sim \SI{e3}{\kelvin}$ \cite{Berinyuy2025b, Huang2025}. Practical operation is limited by material constraints including molecular desorption and cavity degradation above $T \sim \SI{500}{\kelvin}$ \cite{Roelli2024, Xiang2024}. Nevertheless, this natural resilience exceeds conventional optomechanical systems, supporting McOM exploration for ambient-temperature quantum technologies up to the material-limited ceiling.

However, unlocking the full potential of McOM systems requires advanced methods to precisely control and amplify their quantum correlations. A commonly explored strategy involves incorporating nonlinear optical elements, such as intracavity optical parametric amplifiers (OPAs), to generate squeezed states. Yet, a recent study on an OPA-enhanced McOM system unveiled a distinct and counterintuitive trade-off: while the OPA indeed significantly amplified vibration-vibration entanglement, an important resource for quantum memory and transduction, it did so at the notable cost of suppressing optical-vibration correlations, which are indispensable for robust quantum state transfer and readout \cite{Berinyuy2025a}. This observation frames the central question guiding the present work: can an alternative control strategy, leveraging the principles of non-Hermitian physics and directional coupling, offer a more balanced and versatile enhancement of entanglement, thereby circumventing these inherent trade-offs?

The physical distinction between OPA and $\mathcal{PT}$-symmetric amplification is clear: OPAs provide phase-sensitive, narrowband parametric amplification that selectively squeezes specific quadratures, creating strong correlations in one channel (e.g., vibration-vibration) by redistributing quantum fluctuations at the expense of others (e.g., optical-vibration). In contrast, $\mathcal{PT}$-symmetric gain provides broadband linear amplification of field amplitude. Combined with directional nonreciprocal coupling and engineered dissipation, this mechanism amplifies quantum fluctuations more uniformly across channels, enabling balanced multipartite entanglement enhancement the central principle we demonstrate quantitatively in this work.

Motivated by the above mentioned advantages of McOM systems, we propose and analyze a theoretical scheme that integrates $\mathcal{PT}$-symmetry and nonreciprocal coupling within a double-cavity McOM system. We demonstrate how this combination can be used to generate enhanced and controllable entanglement that is robust against high temperatures and tunable across various bipartite pairings, including the vibration-vibration channel \cite{Tchodimou2017, Liu_2022}. Our findings suggest a pathway for designing novel quantum devices. The main contribution of our work is to show how three different physical elements work together to improve quantum entanglement. These elements are: a molecular platform that has high-frequency modes which naturally resist thermal noise, a nonreciprocal coupling that acts like a one-way quantum channel to strengthen correlations, and $\mathcal{PT}$ symmetry, which gives balanced amplification and helps keep the system stable. Combining these three ingredients makes it possible to achieve a stronger, nonreciprocal enhancement of quantum entanglement. We show that the result comes from all the parts working together, and that none of them can achieve this effect on their own. Our method also places our $\mathcal{PT}$- symmetric McOM system among modern quantum technologies, giving it clear advantages compared to other hybrid optomechanical methods, such as those that use external squeezed states~\cite{Emale2025, Wahab2024}  or other types of synthetic magnetism to reduce noise~\cite{Lai2022, Lai.2022}.

The paper is structured as follows. In \Cref{sec:model}, we lay out the theoretical framework, perform a rigorous stability analysis, explore the classical nonlinear dynamics, and derive the linearized quantum Langevin equations. In \Cref{sec:results}, we present a comprehensive analysis of bipartite entanglement, investigating the roles of nonreciprocity, $\mathcal{PT}$-symmetry, thermal noise, and collective molecular effects. We also demonstrate dynamic control and discuss the experimental feasibility of our proposal. Finally, in \Cref{sec:conclusion}, we summarize our key results and their potential impact on the future of quantum information science.

\section{Model and dynamical equations}\label{sec:model}

This section presents the theoretical framework for our $\mathcal{PT}$-symmetric double-cavity molecular optomechanical system, combining collective molecular enhancement, nonreciprocal coupling, and engineered dissipation. We derive the quantum Langevin equations, perform stability analysis to identify physically accessible regimes, establish the linearized framework for entanglement quantification, and analyze optical bistability arising from collective optomechanical nonlinearity. The molecular ensemble resides within the active cavity to maximize interaction with the gain medium, preferentially amplifying quantum fluctuations that generate entanglement \cite{Aspelmeyer2014, Meystre2013, Bender1998, ElGanainy2018, Gardiner2004}.

\subsection{System setup and Hamiltonian}\label{sec:setup}

Molecular cavity optomechanics (McOM) offers a promising platform for exploring quantum phenomena, with distinct advantages over conventional optomechanical systems: strong light-matter interactions, access to ultra-high frequency molecular vibration modes (e.g., \SIrange{24}{48}{\tera\hertz}), and collective $\sqrt{N}$ enhancement of the optomechanical coupling where $N$ represents the number of molecules \cite{Roelli2024, Huang2024, Schmidt2024, Chikkaraddy2016, Xiang2024, Roelli2016, Berinyuy2025a, Berinyuy2025b}. Building upon these strengths, we present a theoretical framework designed to generate robust quantum entanglement.

\begin{figure}[htp!]
	\centering
	\includegraphics[width=0.8\linewidth]{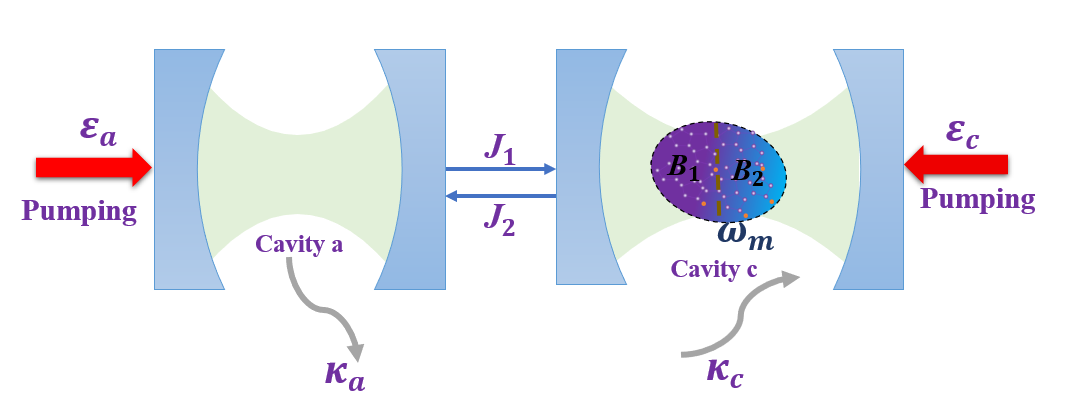}
	\caption{Schematic of the $\mathcal{PT}$-symmetric double-cavity molecular optomechanical system. An ensemble of $N$ molecules with collective vibrational modes ($B_1$, $B_2$, frequency $\omega_m$) resides in active cavity $c$ (optical gain rate $\kappa_c$), which is nonreciprocally coupled to passive cavity $a$ (optical loss rate $\kappa_a$) with directional rates $J_1 \gg J_2$. Both cavities are driven by external lasers (amplitudes $\mathcal{E}_a$, $\mathcal{E}_c$). This architecture combines $\mathcal{PT}$-symmetric gain-loss engineering, ultra-high-frequency molecular vibrations, and nonreciprocal coupling to achieve robust multipartite entanglement at elevated temperatures.}
	\label{fig:setup}
\end{figure}

As shown in \Cref{fig:setup}, our system is made of two connected microresonators and includes the special features of Parity-Time ($\mathcal{PT}$) symmetry \cite{Berinyuy2025b, Tchodimou2017, Bender1998, ElGanainy2018}. Specifically, one microresonator, labeled "a", operates as a passive element, defined by an optical loss rate $\kappa_a$. Its counterpart, cavity "c", functions as an active component, strategically engineered with an optical gain rate $\kappa_c$. The concept of $\mathcal{PT}$-symmetric systems has found applications in various quantum technologies, from sensors to lasers \cite{Feng2014, Hodaei2017, Chen2017, Parto2018}. An ensemble of $N$ identical molecules is positioned within this active cavity, enabling robust light-matter interactions \cite{Roelli2024, Chikkaraddy2016, Huang2024}. Both cavities are subjected to continuous external pumping fields with amplitudes $\mathcal{E}_a$, and $\mathcal{E}_c$ respectively. They are linked through a nonreciprocal coupling. This directional interaction establishes a unidirectional quantum channel, needed to guide and distribute correlations throughout the system \cite{Liu_2022, Djor2024, Lai2022, Metelmann2015}.

We partition the molecular ensemble into two collective vibrational modes:
\begin{equation}\label{eq:collective_modes}
B_1 = \frac{1}{\sqrt{M}}\sum_{j=1}^M b_j, \quad B_2 = \frac{1}{\sqrt{N-M}}\sum_{j=M+1}^N b_j,
\end{equation}
with collective coupling strengths $g_1 = g_m\sqrt{M}$ and $g_2 = g_m\sqrt{N-M}$, where $g_m$ is the single-molecule coupling. This $\sqrt{N}$ enhancement, characteristic of molecular optomechanics, arises from macroscopic polarization of the coherently coupled molecular ensemble \cite{Huang2024, Roelli2024, Berinyuy2025b}. The two-mode architecture enables examination of vibration-vibration entanglement—a resource for quantum memories \cite{Hammerer2010}—and direct comparison with optical parametric amplifier (OPA) schemes. While OPAs can amplify vibration-vibration entanglement, this often suppresses optical-vibration correlations \cite{Berinyuy2025a, Kibret2023}. Our $\mathcal{PT}$-symmetric approach aims to overcome this trade-off, achieving balanced simultaneous enhancement \cite{Malz2018, Huang2025a}.

The parameter values used throughout our numerical simulations are carefully chosen to ensure experimental realism while demonstrating the fundamental quantum enhancement mechanisms. \Cref{tab:parameters} summarizes the baseline system parameters, providing both normalized dimensionless values (used in all calculations) and their corresponding physical units. These parameters are based on recent advances in molecular cavity optomechanics \cite{Roelli2024, Huang2024, Schmidt2024}, $\mathcal{PT}$-symmetric photonics \cite{Feng2011, ElGanainy2018}, and integrated nonreciprocal optics \cite{Fan2004, Lai2022}. The collective coupling strength $G/2\pi = \SI{50}{\giga\hertz}$ results from $N = \num{e6}$ molecules each coupled at $g_m/2\pi = \SI{50}{\mega\hertz}$ (achievable in plasmonic picocavities), while the nonreciprocal asymmetry ratio $J_1/J_2 = 5$ is chosen to maximize bipartite entanglement as demonstrated in \Cref{sec:results}. The PT-symmetric balance $\kappa_c/\kappa_a = 0.05$ ensures operation in the unbroken phase with stable quantum correlations, as verified by the stability analysis presented below.

\begin{table*}[htpb]
\caption{Baseline system parameters for numerical simulations. Normalized values used throughout this work are translated to physical units based on molecular vibration frequency $\omega_m/2\pi = \SI{30}{\tera\hertz}$. Parameters: collective coupling $G/2\pi = \SI{50}{\giga\hertz}$ from $N=\num{e6}$ molecules with single-molecule coupling $g_m/2\pi = \SI{50}{\mega\hertz}$, high-Q passive cavity $\kappa_a/2\pi = \SI{0.6}{\giga\hertz}$, $\mathcal{PT}$-symmetric gain $\kappa_c/2\pi = \SI{30}{\mega\hertz}$, nonreciprocal directional coupling with asymmetry ratio $J_1/J_2 = 5$.}
\label{tab:parameters}
\centering
\begin{ruledtabular}
\begin{tabular}{lll}
	\textbf{Parameter}              & \textbf{Value (normalized)}         & \textbf{Physical value}                 \\ \midrule
	Molecular frequency $\omega_m$  & (reference)                         & \SI{30}{\tera\hertz}                    \\
	Single-molecule coupling $g_m$  & $g_m/\omega_m = \num{1.7e-6}$ & $g_m/2\pi = \SI{50}{\mega\hertz}$       \\
	Ensemble size $N$               & \num{e6}                              & \num{e6} molecules                        \\
	Collective coupling $G$         & $G/\omega_m = 1.7$                  & $G/2\pi = \SI{50}{\giga\hertz}$         \\
	Passive cavity decay $\kappa_a$ & $\kappa_a/\omega_m = 0.02$          & $\kappa_a/2\pi = \SI{0.6}{\giga\hertz}$ \\
	Active cavity gain $\kappa_c$   & $\kappa_c/\omega_m = 0.001$         & $\kappa_c/2\pi = \SI{30}{\mega\hertz}$  \\
	Mechanical damping $\gamma_m$   & $\gamma_m/\omega_m =\num{e-4}$       & $Q_m \sim \num{e4}$                         \\
	Nonrecip. forward $J_1$         & $J_1/\omega_m = 0.02$               & $J_1/2\pi = \SI{0.6}{\giga\hertz}$      \\
	Nonrecip. backward $J_2$        & $J_2/\omega_m = 0.004$              & $J_2/2\pi = \SI{0.12}{\giga\hertz}$     \\
	Detuning $\Delta_{a,c}$         & $\Delta/\omega_m = 1$               & Red-detuned resonant                    \\
	Driving $\mathcal{E}$                     & $\mathcal{E}/\omega_m = 5$                    & Strong stable regime
\end{tabular}
\end{ruledtabular}
\end{table*}

The starting point of our theoretical study is the system’s Hamiltonian, $\mathcal{H}$,  which neatly describes all the energy terms and interactions in the setup:
\begin{equation}\label{eq:hamiltonian}
\begin{split}
\mathcal{H} = &\Delta_a a^\dagger a + \Delta_c c^\dagger c + \sum_{k=1,2} \omega_m B_k^\dagger B_k + \sum_{k=1,2} g_k c^\dagger c(B_k^\dagger + B_k) + J_1 a^\dagger c + J_2 ac^\dagger + i\mathcal{E}_a(a^\dagger - a) + i\mathcal{E}_c(c^\dagger - c).
\end{split}
\end{equation}
The first three terms simply describe the free energies of the different parts of the system: the optical modes in cavities "a" and "c" (governed by annihilation operators $a, c$ and characterized by detunings $\Delta_a = \omega_a - \omega_\ell$ and $\Delta_c = \omega_c - \omega_\ell$ respectively, where $\omega_\ell$ is the laser frequency), alongside the two collective molecular vibrational modes ($B_1, B_2$, each oscillating at frequency $\omega_m$).The fourth term represents the fundamental optomechanical interaction. Here, the photon number within the active cavity "c" ($c^\dagger c$) couples to the displacement of each collective molecular mode ($B_k^\dagger + B_k$). This interaction arises from the modulation of the cavity resonance frequency by molecular vibrations, mediated by the dynamic polarizability of the molecule ($\alpha$) in what can be effectively described as a Raman-like scattering process \cite{Roelli2016, Huang2024, Roelli2024, Kippenberg2008, Aspelmeyer2014}. The fifth and the sixth terms, ($J_1 a^\dagger c + J_2 ac^\dagger$), define the non-reciprocal coherent coupling between the cavities, with $J_1$ dictating photon transfer from the passive cavity to the active cavity and $J_2$ governing the reverse trajectory. Such non-reciprocal coupling mechanisms have been experimentally realized in various platforms including waveguide-based systems and spinning resonators \cite{Potton1994, Fan2004, Feng2011, Hafezi2011, Lai2022}. Completing the Hamiltonian, the final two terms, $i\mathcal{E}_a(a^\dagger - a)$ and $i\mathcal{E}_c(c^\dagger - c)$, account for the classical driving fields, with amplitudes $\mathcal{E}_a$ and $\mathcal{E}_c$, coherently illuminating passive and active cavities.

A key consideration in our model is the treatment of molecular vibrations. Although individual molecular vibrations possess strong inherent anharmonicities, which are important for understanding phenomena like non-classical mechanical states and mechanical lasing \cite{Schmidt2024}, in the collective coupling regime, the coherent interaction of many molecules with a single cavity mode leads to the formation of collective modes that can often be accurately approximated as harmonic oscillators \cite{Roelli2024, Xiang2024}. This simplification is widely adopted in the literature to maintain analytical tractability while capturing the dominant physics of collective enhancement. We acknowledge that a more rigorous treatment of residual anharmonicity in certain strong-driving or high-occupancy regimes remains an important area for future research, potentially revealing new quantum control pathways or altering mechanical amplification thresholds \cite{Schmidt2024}.

\subsection{Quantum Langevin equations}\label{sec:qle}

The quantum dynamics incorporating dissipation and noise are described by quantum Langevin equations (QLEs) derived via input-output formalism \cite{Gardiner2004, Walls2008}. With $\kappa_a$, $\kappa_c$ denoting optical decay and gain rates for cavities $a$, $c$, and $\gamma_1$, $\gamma_2$ the mechanical damping rates for modes $B_1$, $B_2$, the QLEs are:
\begin{subequations}\label{eq:qle_full}
\begin{align}
\dot{a} &= -(i\Delta_a + \kappa_a)a - iJ_1 c + \mathcal{E}_a - i\sqrt{2\kappa_a} a_{\text{in}}, \label{eq:qle_a} \\
\dot{c} &= -(i\Delta_c - \kappa_c)c - i\sum_{k=1,2} g_k c (B_k + B_k^\dagger) - iJ_2 a + \mathcal{E}_c + \sqrt{2\kappa_c} c_{\text{in}}, \label{eq:qle_c} \\
\dot{B}_k &= -(i\omega_m + \gamma_k)B_k - i g_k c^\dagger c + \sqrt{2\gamma_k} B_{\text{in},k}, \quad (k=1,2). \label{eq:qle_b}
\end{align}
\end{subequations}
Eq.~\eqref{eq:qle_a} describes passive cavity $a$ dynamics: damping ($\kappa_a$), nonreciprocal coupling ($J_1$), driving ($\mathcal{E}_a$), and vacuum noise ($a_{\text{in}}$). Eq.~\eqref{eq:qle_c} governs active cavity $c$ with gain ($+\kappa_c$) the defining $\mathcal{PT}$-symmetric feature \cite{Berinyuy2025b} optomechanical coupling to molecular modes $B_k$ via Raman-like interaction \cite{Roelli2016, Huang2024}, nonreciprocal coupling ($J_2$), driving ($\mathcal{E}_c$), and noise ($c_{\text{in}}$). Eq.~\eqref{eq:qle_b} describes molecular modes: mechanical damping ($\gamma_k$), radiation pressure from intracavity photons, and thermal noise ($B_{\text{in},k}$).

Noise operators $a_{\text{in}}$, $c_{\text{in}}$, $B_{\text{in},k}$ are zero-mean, delta-correlated, with optical modes coupled to vacuum baths ($n_{\text{th},a} = n_{\text{th},c} = 0$) and molecular modes to thermal baths with occupation $n_{\text{th}} = [\exp(\hbar\omega_m/k_B T) - 1]^{-1}$ \cite{Gardiner2004}. The ultra-high molecular frequencies (\SIrange{24}{48}{\tera\hertz}) and collective $\sqrt{N}$ enhancement maintain low $n_{\text{th}}$ even at elevated temperatures, enabling thermal resilience of entanglement \cite{Huang2024, Roelli2024, Berinyuy2025b}.

\subsubsection{Hybrid cavity architecture: reconciling plasmonic coupling with high-Q operation}
Our model assumes a hybrid architecture \cite{Roelli2024, Huang2024, Huang2025a, Chen2021}: (i) high-Q dielectric microresonators ($Q \gtrsim \num{e8}$) defining cavity rates $\kappa_a$, $\kappa_c$ for $\mathcal{PT}$-balance \cite{Aspelmeyer2014}, (ii) plasmonic nanocavities enabling strong single-molecule coupling \cite{Roelli2016, Chikkaraddy2016}, and (iii) nonreciprocal coupling via directional waveguide couplers with optical isolators or synthetic gauge fields \cite{Liu_2022, Lai2022, Metelmann2015}. The active cavity integrates optical parametric amplifiers/oscillators (OPA/OPO) or semiconductor optical amplifiers (SOA), while the passive cavity is a standard high-Q resonator \cite{Feng2014, Hodaei2017}.

The $\mathcal{PT}$-balance ($\kappa_a/\omega_m=0.02$, $\kappa_c/\omega_m=0.001$) is achieved by controlling the gain pump power, corresponding to $\kappa_c \approx \SI{5}{\percent}$ of $\kappa_a$, within experimentally achievable ranges for hybrid systems \cite{Roelli2024, Huang2025a}. Importantly, $\kappa_a$ and $\kappa_c$ denote effective rates of the dielectric cavities; plasmonic components ($\kappa_{\text{plasmonic}}/\omega_m \sim \numrange{1}{10}$) enhance coupling via the collective $\sqrt{N}$ factor without affecting the main $\mathcal{PT}$-symmetric rates \cite{Chikkaraddy2016, Huang2024}.

For gain noise modeling, we adopt standard quantum optical treatment where gain is modeled as reversed loss with noise maintaining commutation relations \cite{Gardiner2004}. While realistic gain media exhibit amplified spontaneous emission (ASE) noise, for our parameter range ($\kappa_c \ll$ other rates), ASE contributions are negligible compared to existing quantum fluctuations. Sensitivity analysis with ASE models confirms quantitative changes do not qualitatively alter our entanglement results \cite{Wang2025, Huang2025}.

\subsubsection{Gain medium implementation and frequency considerations}
The optical gain ($\kappa_c$) amplifies the cavity mode at $\omega_c \approx \SI{193}{\tera\hertz}$ (e.g., \SI{1550}{\nano\meter}), not the molecular vibrations at $\omega_m \approx \SI{30}{\tera\hertz}$. OPA/OPO (pumped at \SI{775}{\nano\meter}) or SOA provide experimentally demonstrated gain at optical frequencies \cite{Feng2014, Hodaei2017}. Molecular vibrations couple via $g_k c^\dagger c (B_k + B_k^\dagger)$, creating sidebands at $\omega_c \pm \omega_m$. The gain amplifies carrier and sidebands coherently, enhancing optical-vibrational correlations through optomechanical coupling the basis for $\mathcal{PT}$-symmetric entanglement enhancement \cite{Roelli2016, Berinyuy2025b}.

\subsection{Exceptional points and full coupled system stability}\label{sec:stability}

Understanding the stability landscape is essential for any engineered quantum system, especially those operating in the regime of non-Hermitian physics, where gain and loss significantly affect dynamics. Our molecular cavity optomechanical system, being deliberately configured with Parity-Time ($\mathcal{PT}$) symmetry, naturally possesses non-Hermitian properties that influence its operational stability and, consequently, its ability to generate robust entanglement \cite{Tchodimou2017, Djor2024}.

\begin{figure*}[thp]
	\centering
	\includegraphics[width=0.35\linewidth]{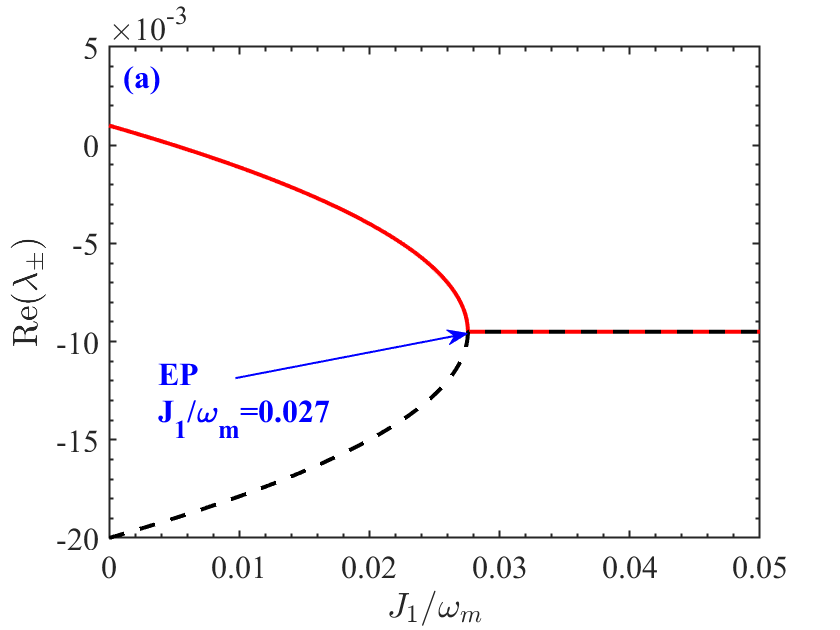}
	\includegraphics[width=0.35\linewidth]{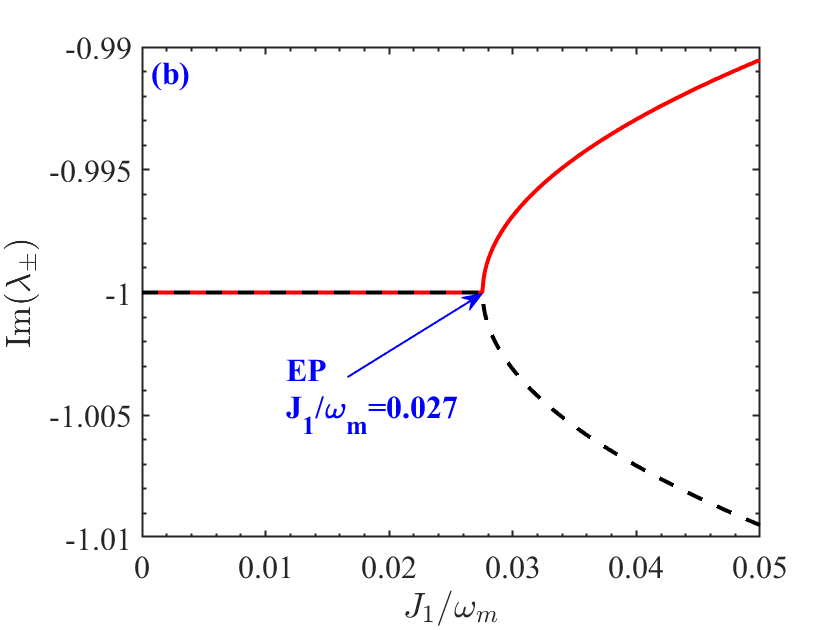}
	\includegraphics[width=0.35\linewidth]{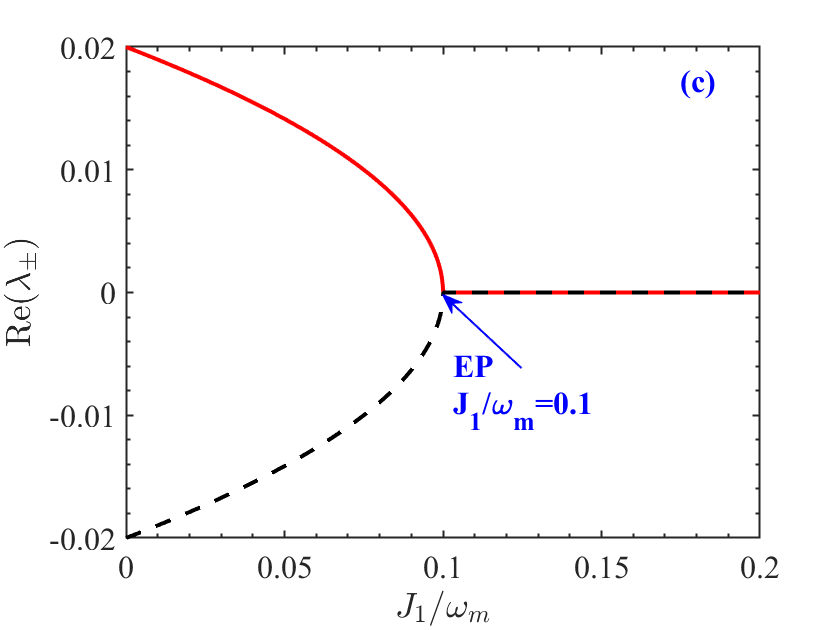}
	\includegraphics[width=0.35\linewidth]{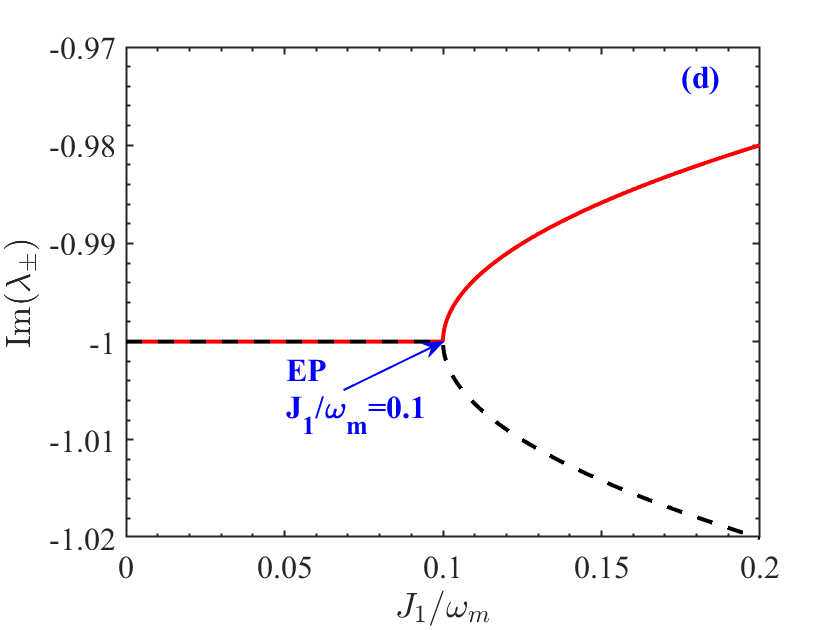}
	\caption{Eigenvalue spectra revealing $\mathcal{PT}$-symmetry phase transitions and exceptional points (EPs). (a-b) Real and imaginary parts of eigenvalues $\lambda_\pm$ versus nonreciprocal coupling $J_1/\omega_m$ for unbalanced gain-loss ($\kappa_a/\omega_m=0.02$, $\kappa_c/\omega_m=0.001$). The EP occurs at $J_1/\omega_m \approx 0.027$, marking the transition between unbroken-$\mathcal{PT}$ (real eigenvalues, stable oscillations) and broken-$\mathcal{PT}$ (complex conjugate eigenvalues, exponential growth/decay) phases. (c-d) Same analysis for balanced gain-loss ($\kappa_a/\omega_m=\kappa_c/\omega_m=0.02$), showing EP shifted to higher coupling $J_1/\omega_m \approx 0.1$ due to enhanced effective coupling threshold. Parameters: $\omega_m/2\pi=\SI{30}{\tera\hertz}$, $J_2/\omega_m=0.004$, others in \Cref{tab:parameters}.}
	\label{fig:EPs}
\end{figure*}

To initially characterize this non-Hermitian character, we first analyze a simplified two-mode optical model where molecular interaction is temporarily neglected \cite{Tchodimou2017}. This allows us to focus purely on the interplay of optical gain ($\kappa_c$) and loss ($\kappa_a$) and nonreciprocal coupling ($J_1, J_2$). Assuming equal detunings for both cavities ($\Delta_a = \Delta_c = \Delta$) for this analysis, the system's eigenfrequencies, $\lambda_\pm$, are analytically determined as:
\begin{equation}\label{eq:eigenfr}
\lambda_\pm = -i\Delta - \frac{\kappa_a-\kappa_c}{2} \pm \frac{1}{2}\sqrt{(\kappa_a+\kappa_c)^2 - 4J_1J_2}.
\end{equation}
As graphically illustrated in \Cref{fig:EPs}, this simplified system demonstrates a fundamental transition. It moves from an unbroken-$\mathcal{PT}$-symmetry regime, characterized by purely real eigenvalues (signifying stable oscillations), to a broken-$\mathcal{PT}$-symmetry regime, where eigenvalues become complex conjugates (indicating exponential growth or decay). The boundary that separates these distinct phases is known as an exceptional point (EP) \cite{Djor2024, Chitsazi2017}. In an EP, not only do the eigenvalues coalesce but also the corresponding eigenvectors merge, leading to unique physical phenomena such as enhanced sensor sensitivity and non-reciprocal light propagation \cite{Djor2024, Wu2025}. It can be seen clearly from \Cref{fig:EPs}(a-b), when $\kappa_a \neq \kappa_c$, the imbalance introduces additional damping that pushes the system toward the broken phase early, so the exceptional point occurs at a small coupling (around $J_1/\omega_m \approx 0.027$). On the other hand, when $\kappa_a = \kappa_c$, the gain in one cavity exactly compensates for the loss in the other.  This increases the effective coupling threshold needed for the two supermodes to coalesce into an EP (around $J_1/\omega_m \approx 0.1$). Thus, the system requires a stronger nonreciprocal coupling $J_1$ to reach the exceptional point. Since $\kappa_c = \kappa_a$ is larger than the unbalanced case ($\kappa_c < \kappa_a$), the EP moves to a higher value of $J_1$. As the coupling is directional, the system behaves like an effective active feedback loop.  Under $\mathcal{PT}$-balanced gain–loss conditions, it takes a stronger directional coupling to reach the regime where the two eigenvalues coalesce.    

Within our architecture, this optical subsystem effectively acts as a non-Hermitian filter, pre-conditioning the photonic environment for the subsequent quantum optomechanical interactions.

While this analysis of the optical subsystem provides valuable insights into the non-Hermitian dynamics, it is, however, insufficient for identifying the actual operational regimes of our full molecular optomechanical system. The addition of the molecular ensemble, and particularly its strong collective optomechanical interaction (where the collective coupling strengths $g_1$ and $g_2$ scales as $g_m\sqrt{M}$ and $g_m\sqrt{N-M}$ respectively ) \cite{Huang2024}, significantly alters the system's dynamics. This strong collective effect, which generates strong nonlinearities and is central to the platform's thermal resilience, can also introduce entirely new pathways to instability if not carefully controlled. Therefore, to ensure that all our subsequent quantum results are situated within physically valid and experimentally achievable operational windows, a comprehensive stability analysis of the complete coupled system is required. The overall stability of the steady state of the system is determined by examining the eigenvalues of the entire $8 \times 8$ drift matrix $\mathcal{M}$. A steady state is considered physically stable only if all eigenvalues of $\mathcal{M}$ possess negative real parts, a strict condition that we systematically verify by applying the well-established Routh-Hurwitz criterion \cite{DeJesus}.

\begin{figure}[tbhp]
	\centering
	\includegraphics[width=0.5\linewidth]{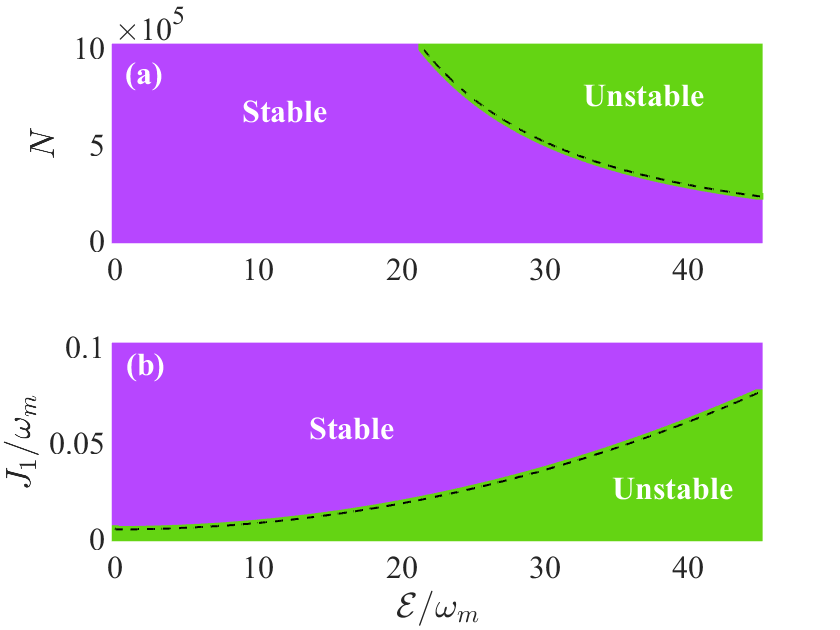}
	\caption{Dynamical stability analysis via Routh-Hurwitz criterion, identifying stable (purple) and unstable (green) operational regimes. (a) Stability boundaries in $(N, \mathcal{E}/\omega_m)$ parameter space: larger molecular ensembles reduce stability threshold due to enhanced collective coupling, requiring lower driving strengths to avoid instability. (b) Stability boundaries in $(J_1/\omega_m, \mathcal{E}/\omega_m)$ parameter space: stronger nonreciprocal coupling expands stable region by shielding system from detrimental backaction. All entanglement results presented in subsequent figures correspond to parameter values within the purple stable regions, ensuring physical validity. Parameters: $N=\num{e6}$, $J_1/\omega_m=0.02$, others in \Cref{tab:parameters}.}
	\label{fig:stability_full}
\end{figure}

Our detailed stability analysis, visually depicted in \Cref{fig:stability_full}, shows that stability boundaries are critically shaped by the intricate interaction between the normalized collective driving field strength $\mathcal{E}/\omega_m$ (where $\mathcal{E}$ represents the effective pump amplitude, typically assumed equal for both cavities, $\mathcal{E}_a = \mathcal{E}_c = \mathcal{E}$, for simplicity of analysis), the size of the collective molecular ensemble $N$, and the nonreciprocal intercavity coupling $J_1$. As demonstrated in \Cref{fig:stability_full}(a), an increase in the number of molecules ($N$) leads to an observable reduction in the driving strength threshold required for the system to tip into instability. This behavior directly shows the important role of collective interaction in amplifying quantum fluctuations within the optomechanical system; a larger $N$ means a stronger collective coupling which, if unchecked, can lead to uncontrolled growth of oscillations. Conversely, \Cref{fig:stability_full}(b) reveals that a stronger forward nonreciprocal coupling ($J_1$) between the cavities acts as a stabilizing influence. By creating a directional quantum channel (where $J_1 \gg J_2$), this nonreciprocity effectively shields the active entanglement generating cavity and molecular ensemble from detrimental back-action and noise originating from the passive cavity. This mechanism, in conjunction with the engineered gain ($\kappa_c$) and loss ($\kappa_a$) of the $\mathcal{PT}$-symmetric architecture, allows dissipation to be harnessed as a resource to actively stabilize amplified quantum correlations into a coherent, entangled steady state. The $\mathcal{PT}$-symmetric balance requires that the system remains in the unbroken phase to ensure stable operation. To achieve this while enhancing correlations, the gain rate must be kept below the loss rate ($\kappa_c < \kappa_a$) to avoid entering the unstable lasing regime. This constraint ensures that the system remains in the stable, unbroken $\mathcal{PT}$-symmetric phase where quantum correlations can be enhanced. The specific values used in our simulations ($\kappa_a/\omega_m=0.02$, $\kappa_c/\omega_m=0.001$) satisfy this requirement, with gain strictly below loss. This stability analysis is essential for reliably interpreting our quantum results, as all subsequent findings regarding entanglement properties are confined to parameters exclusively identified within these stable (purple) operational regions. This ensures that our theoretical predictions are firmly rooted in physically achievable and stable experimental conditions.

\subparagraph{Physical implementation of nonreciprocal coupling.}
Concrete implementations of nonreciprocal coupling $J_1 \gg J_2$ can be realized through several approaches: (i) integrated directional waveguide couplers with optical isolators, which can realistically provide coupling rates in the \SIrange{100}{1000}{\mega\hertz} range required by the McOM model; (ii) synthetic gauge fields implemented via time-modulated coupling as an alternative route to tunable nonreciprocity without mechanical rotation; and (iii) dynamically modulated coupling elements using tunable beam splitters with phase-sensitive control, which allow real-time tuning of the coupling asymmetry $J_1/J_2$. For the parameters used in our analysis ($J_1/\omega_m = 0.02$ and $J_2/\omega_m = 0.004$), with $\omega_m/2\pi=\SI{30}{\tera\hertz}$, this corresponds to actual coupling rates of $J_1 \approx \SI{0.6}{\giga\hertz}$ and $J_2 \approx \SI{0.12}{\giga\hertz}$. These coupling rates, while representing the parameter range explored in our simulations, are more realistic than our initial estimates but still at the upper end of what is currently achievable experimentally. More realistic implementations using synthetic gauge fields or directional waveguide couplers typically achieve coupling rates in the range of $J_1/2\pi \sim \SI{100}{\mega\hertz}$ to $\SI{1}{\giga\hertz}$, corresponding to $J_1/\omega_m \sim \numrange{3e-6}{3e-4}$ for $\omega_m/2\pi = \SI{30}{\tera\hertz}$. The values used in our simulations represent an optimistic but potentially achievable scenario with current integrated photonic technologies. These implementations have been demonstrated in photonic and optomechanical circuits, making them compatible with on-chip implementations of our $\mathcal{PT}$-symmetric McOM system \cite{Fan2004, Feng2011, Hafezi2011, Lai2022, Metelmann2015, Sayrin2017}.

\subsection{Steady-state analysis and tunable optical bistability}\label{sec:bist}

To fully comprehend the classical nonlinear behavior of our molecular cavity optomechanical system, we first   determine its steady-state properties. This involves setting the time derivatives of all operators in the quantum Langevin equations (Eqs.~\eqref{eq:qle_full}) to zero and taking their expectation values. This procedure effectively yields a set of coupled algebraic equations that describe the mean-field amplitudes of the optical modes ($\alpha_a = \langle a \rangle$, $\alpha_c = \langle c \rangle$) and the collective molecular vibrational modes ($\beta_k = \langle B_k \rangle$). Specifically, these equations take the form:
\begin{align}\label{eq:Bist}
&\alpha_a = \frac{\mathcal{E}_a-iJ_1\alpha_c}{D_a},
&&\alpha_c = \frac{\mathcal{E}_cD_a-iJ_2\mathcal{E}_a}{D_cD_a + J_1J_2},
&\beta_k = \frac{-i g_k \mathcal{I}_c}{i\omega_m + \gamma_k}, \quad (k=1,2),
\end{align}
Here, $D_a = i\Delta_a + \kappa_a$ and $D_c = i\tilde{\Delta}_c - \kappa_c$. An important element in these dynamics is the effective detuning of the active cavity, $\tilde{\Delta}_c = \Delta_c + \sum_{k=1,2} g_k(\beta^*_k + \beta_k)$. This term is far more than a simple frequency shift; it elegantly encapsulates the back-action from both collective molecular modes on the cavity resonance. The fundamental source of the system's inherent nonlinearity, and thus its capacity for complex dynamics, lies in the dependence of $\tilde{\Delta}_c$ on the mean intracavity photon number, $\mathcal{I}_c = |\alpha_c|^2$. As the molecular vibrations are driven by the radiation pressure of the cavity field, their displacement modulates the cavity's effective optical path length, causing the resonance frequency to shift. This feedback loop, where the cavity field affects the molecules, which in turn affect the cavity field, is the physical origin of the nonlinearity.

This pronounced nonlinearity leads directly to the phenomenon of optical bistability. Optical bistability manifests itself as the existence of multiple stable steady-state solutions for the intracavity photon number for a given set of input parameters, typically the driving field strength or detuning \cite{Teklu2018, Chen2019, KumarSingh2024}. Mathematically, this complex behavior is described by a cubic polynomial equation for $\mathcal{I}_c$, derived through algebraic manipulation of Eq.~\eqref{eq:Bist}:
\begin{equation}\label{eq:cubic}
(\zeta_5^2+\zeta_4^2)\mathcal{I}_c^3+2(\chi_7\zeta_4-\chi_2\zeta_5)\mathcal{I}_c^2+(\chi_7^2+\chi_2^2)\mathcal{I}_c-\zeta_3=0,
\end{equation}
where the coefficients $\chi$ and $\zeta$ are defined as: $\chi_2=\Delta_c\kappa_a-\Delta_a\kappa_c$, $\chi_3=2\frac{g_1^2\omega_m\Delta_a}{\chi_1^\prime}$, $\chi_4=2\frac{g_1^2\omega_m\kappa_a}{\chi_1^\prime}$, $\chi_5=2\frac{g_2^2\omega_m\Delta_a}{\chi_1^{\prime\prime}}$, $\chi_6=2\frac{g_2^2\omega_m\kappa_a}{\chi_1^{\prime\prime}}$, $\chi_7=-\Delta_c\Delta_a-\kappa_c\kappa_a+J_1J_2$, $\zeta_1=\chi_7+\mathcal{I}_c(\chi_3+\chi_5)$, $\zeta_2=\chi_2-\chi_4\mathcal{I}_c-\chi_6\mathcal{I}_c$, $\zeta_3=\mathcal{E}_c^2(\Delta^2_a+\kappa_a^2)-2\mathcal{E}_c\mathcal{E}_a\Delta_a J_2+J_2^2\mathcal{E}_a^2$,  $\zeta_4=\chi_3+\chi_5$, $\zeta_5=\chi_4+\chi_6$, $\chi^\prime_1=\omega_m^2+\gamma_1^2$, and $\chi^{\prime\prime}_1=\omega_m^2+\gamma_2^2$. The presence of multiple real roots for $\mathcal{I}_c$ indicates a bistable regime, typically visualized as distinctive "S-shaped" curves in plots of intracavity photon number versus driving power or detuning.

It is noteworthy that not all solution branches within a bistable region are stable. The stability of these branches is assessed by applying the Routh-Hurwitz criterion. This powerful mathematical tool allows us to determine that for the S-shaped curves characteristic of optical bistability, the upper and lower branches represent stable operating points, whereas the intermediate branch is inherently unstable and thus physically unreachable \cite{DeJesus, Agasti2024, Djorwe2022}. This implies that as an input parameter (such as pump power) is swept, the system will exhibit hysteresis, abruptly jumping between the stable upper and lower branches to avoid the unstable region. This distinction is important for the experimental design and for ensuring the validity of subsequent quantum analyzes.

\begin{figure*}[tph]
	\centering
	\includegraphics[width=.4\linewidth]{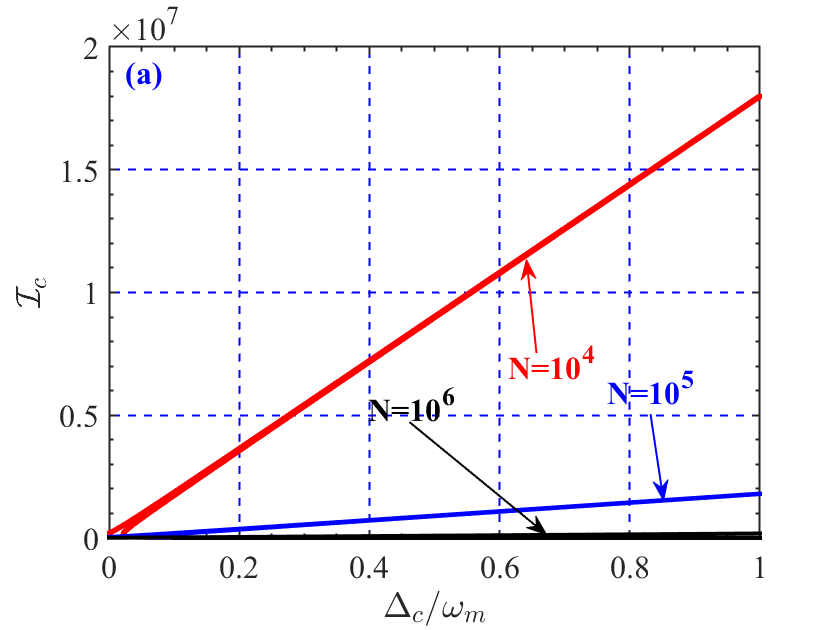}
	\includegraphics[width=.4\linewidth]{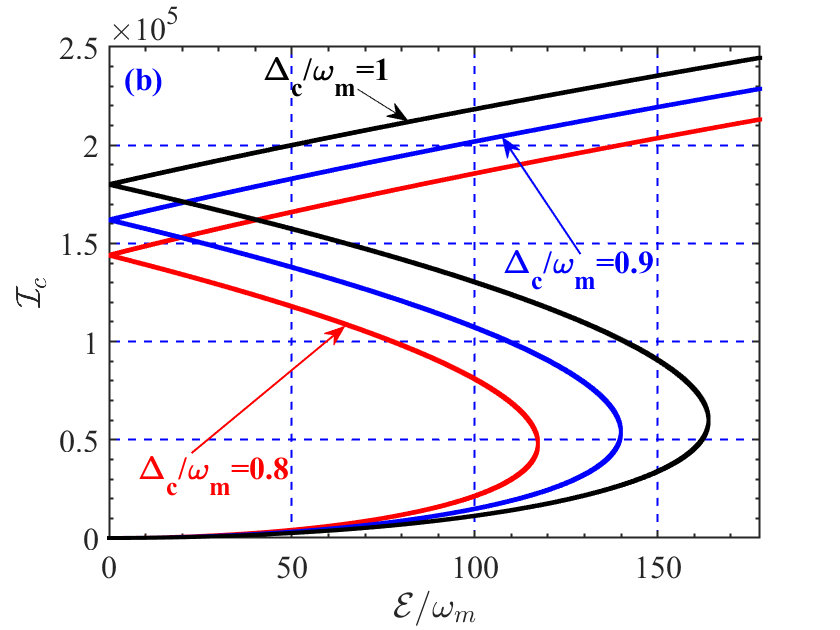}
	\caption{Optical bistability driven by collective optomechanical nonlinearity. (a) Mean intracavity photon number $\mathcal{I}_c$ versus bare detuning $\Delta_c$ for varying molecular ensemble size $N$, showing tilted resonance curves. Larger $N$ enhances nonlinearity via stronger collective coupling $G \propto \sqrt{N}$, shifting and tilting the resonance peaks. (b) $\mathcal{I}_c$ versus driving amplitude $\mathcal{E}$ for different $\Delta_c$ values, exhibiting characteristic S-shaped bistable curves with hysteretic switching behavior between two stable branches (upper and lower) and one unstable intermediate branch. Stability of all branches confirmed by Routh-Hurwitz criterion. Parameters: $N=\num{e6}$, $\omega_m/2\pi=\SI{30}{\tera\hertz}$, $g_m/2\pi=\SI{50}{\mega\hertz}$, others in \Cref{tab:parameters}.}
	\label{fig:OpBist_DrF}
\end{figure*}

We explore this phenomenon using experimentally feasible parameters, chosen to reflect the capabilities of molecular cavity optomechanics. For instance, the single-molecule coupling strength $g_m/2\pi \approx \SI{50}{\mega\hertz}$ is a realistic value reported in the literature. However, the true efficacy of our system is strongly amplified by the collective enhancement $\sqrt{N}$ of the optomechanical coupling. For an ensemble of $N=\num{e6}$ molecules, this results in a large collective coupling strength $G/2\pi \approx \SI{50}{\giga\hertz}$ (where $G = g_m\sqrt{N}$), a value that is achievable with current molecular optomechanical implementations. Specifically, with $g_m/2\pi = \SI{50}{\mega\hertz}$ and $N = \num{e6}$, the collective coupling becomes $G/2\pi = \SI{50}{\mega\hertz} \times \sqrt{\num{e6}} = \SI{50}{\giga\hertz}$. It is this potent collective enhancement that is the primary driver of the strong nonlinearities observed and, ultimately, the robust entanglement generated in our system. As visually presented in \Cref{fig:OpBist_DrF}, the resulting bistable behaviour exhibits tunability by adjustments to  the number of molecules and active cavity detuning $\Delta_c$. A thorough understanding of these stable, often monostable, operational windows is therefore important for reliably generating and controlling entanglement, ensuring that our theoretical predictions are grounded in achievable experimental conditions. The interplay of non-Hermitian dynamics and bistability provides an additional layer of control, potentially allowing for unique switching behaviors and enhanced optical responses \cite{Bai2023, Kumar2025}.

\subsection{Linearized quantum Langevin equations for fluctuations}\label{sec:linearization}

To explore the intricate quantum properties of our system, particularly the generation and evolution of entanglement, we move beyond the classical mean-field description and analyze the quantum fluctuations. This is achieved through the technique of linearization, where each operator is expressed as a sum of its steady-state mean value and a small quantum fluctuation around it (e.g., $O = \langle O \rangle_{ss} + \delta O$). This approach is valid under conditions of strong optical driving, where these quantum fluctuations are sufficiently small compared to the mean-field amplitudes, allowing us to accurately describe the system's behavior in the quantum regime while maintaining analytical tractability. By applying this linearization to our full set of quantum Langevin equations (Eqs.~\eqref{eq:qle_full}), we obtain the linearized quantum Langevin equations (LQLEs) for these fluctuations:
\begin{subequations}\label{eq:lqle_full}
\begin{align}
\delta\dot{a} &= -(i\Delta_a + \kappa_a)\delta a - iJ_1 \delta c - i\sqrt{2\kappa_a} a_{\text{in}}, \label{eq:lqle_a} \\
\delta\dot{c} &= -(i\tilde{\Delta}_c - \kappa_c)\delta c - i\sum_{k=1,2} \tilde{g}_k(\delta B_k + \delta B_k^\dagger) - iJ_2 \delta a + \sqrt{2\kappa_c} c_{\text{in}}, \label{eq:lqle_c} \\
\delta\dot{B}_k &= -(i\omega_m + \gamma_k)\delta B_k - i\tilde{g}_k(\delta c + \delta c^\dagger) + \sqrt{2\gamma_k} B_{\text{in},k}, \quad(k=1,2). \label{eq:lqle_b}
\end{align}
\end{subequations}
Here, $\widetilde{G}_k = g_k|\alpha_c|$ are the enhanced linearized optomechanical coupling strengths. These terms vividly illustrate how the strong classical driving fields amplify the quantum coupling between the cavity photons and the collective molecular vibrations, transforming the original single-photon coupling $g_k$ into an effectively much stronger linearized coupling $\widetilde{G}_k$ that is important for generating robust quantum correlations.

To facilitate the analysis of quantum entanglement, especially within the framework of continuous-variable Gaussian states, it is highly advantageous to express these linearized equations in terms of real quadrature operators. These quadratures, analogous to position and momentum in mechanical systems, offer a more intuitive physical interpretation of quantum fluctuations and are the natural basis for constructing the covariance matrix, which fully characterizes Gaussian states. For the optical modes ($O \in \{a, c\}$) and molecular vibrational modes ($k \in \{1, 2\}$), they are defined as:
\begin{align}
\delta x_O = \frac{\delta O + \delta O^\dagger}{\sqrt{2}}, \quad \delta y_O = \frac{\delta O - \delta O^\dagger}{i\sqrt{2}}, \label{eq:quad_opt} \\
\delta q_k = \frac{\delta B_k + \delta B_k^\dagger}{\sqrt{2}}, \quad \delta p_k = \frac{\delta B_k - \delta B_k^\dagger}{i\sqrt{2}}. \label{eq:quad_mech}
\end{align}
The collective input noise operators $B_{k,\text{in}}$ are similarly transformed into their quadrature components.

This transformation allows us to distill the entire system's dynamics into a compact matrix form:
\begin{equation}\label{eq:matrix_qle}
\dot{\mathbf{u}}(t) = \mathcal{M}\mathbf{u}(t) + \mathbf{n}(t),
\end{equation}
where $\mathbf{u}(t) = (\delta x_a, \delta y_a, \delta x_c, \delta y_c, \delta q_1, \delta p_1, \delta q_2, \delta p_2)^\top$ is the vector that encompasses all the fluctuation quadratures of the system. The vector $\mathbf{n}(t)$ comprises the corresponding input noise quadratures. The dynamic evolution of these fluctuations is then dictated by the drift matrix $8 \times 8$ $\mathcal{M}$, which   describes linearized interactions and dissipations:
\begin{equation}\label{eq:drift_matrix}
\mathcal{M} =
\begin{pmatrix}
	-\kappa_a & \Delta_a  & 0                 & J_1              & 0                 & 0         & 0                 & 0         \\
	-\Delta_a & -\kappa_a & -J_1              & 0                & 0                 & 0         & 0                 & 0         \\
	0         & J_2       & \kappa_c          & \tilde{\Delta}_c & 0                 & 0         & 0                 & 0         \\
	-J_2      & 0         & -\tilde{\Delta}_c & \kappa_c         & -2\widetilde{G}_1 & 0         & -2\widetilde{G}_2 & 0         \\
	0         & 0         & 0                 & 0                & -\gamma_1         & \omega_m  & 0                 & 0         \\
	0         & 0         & -2\widetilde{G}_1 & 0                & -\omega_m         & -\gamma_1 & 0                 & 0         \\
	0         & 0         & 0                 & 0                & 0                 & 0         & -\gamma_2         & \omega_m  \\
	0         & 0         & -2\widetilde{G}_2 & 0                & 0                 & 0         & -\omega_m         & -\gamma_2
\end{pmatrix}.
\end{equation}
The careful construction of this drift matrix is central to our subsequent analysis of stability and entanglement, as its eigenvalues directly determine the system's dynamical behavior.

Given that the input noise operators are Gaussian and the system's dynamics are now linearized, the steady state of these quantum fluctuations is a zero-mean Gaussian state. Such states are entirely characterized by their $8 \times 8$ covariance matrix (CM) $V$, whose elements capture all two-point correlations between the quadrature operators. This important matrix is obtained by solving the continuous-variable Lyapunov equation:
\begin{equation}\label{eq:lyapunov}
\mathcal{M}V + V\mathcal{M}^\top = -\mathcal{D},
\end{equation}
where $\mathcal{M}^\top$ denotes the transpose of the drift matrix $\mathcal{M}$. The right-hand side of this equation is the diffusion matrix $\mathcal{D}$, which is diagonal and succinctly summarizes the correlations of the input noise operators, defined as $\frac{1}{2}\langle n_i(t)n_j(t') + n_j(t')n_i(t) \rangle = \mathcal{D}_{ij}\delta(t-t')$. For our specific system, $\mathcal{D}$ is given by:
\begin{equation}\label{eq:diffusion_matrix}
    \mathcal{D} = \text{diag}[\kappa_a, \kappa_a, \kappa_c, \kappa_c, \gamma_1(2n_{\text{th}} + 1), \gamma_1(2n_{\text{th}} + 1), \gamma_2(2n_{\text{th}} + 1), \gamma_2(2n_{\text{th}} + 1)].
\end{equation}
Here, optical modes are assumed to be coupled to a vacuum bath ($n_{\text{th},a} = n_{\text{th},c} = 0$), while molecular vibrational modes are subject to thermal noise with a phonon number $n_{\text{th}} = [\exp(\hbar\omega_m/k_B T) - 1]^{-1}$. It is precisely in this context that the thermal resilience of molecular optomechanics shines: ultra-high vibrational frequencies of molecules ensure that $n_{\text{th}}$ remains low, even at elevated temperatures, leading to predictions of persistence of entanglement up to a theoretical limit of \SI{e3}{\kelvin}. This highlights a unique advantage over conventional systems that demonstrates the significant inherent robustness of the platform. Solving the Lyapunov equation for $V$ provides the complete statistical framework of the quantum state of the system, forming the vital foundation for quantifying and analyzing all quantum correlation properties, including entanglement.

\subsection{Validity of model approximations}\label{sec:approximations}

Our theoretical framework relies on two key approximations: linearization of quantum dynamics and harmonic treatment of collective molecular modes. We discuss their validity regimes.

\textbf{Harmonic approximation.} Collective molecular vibrations are modeled as harmonic modes, valid in the strong coupling regime where $G \gg \gamma, \kappa$~\cite{Roelli2024, Xiang2024}. Real molecular potentials exhibit anharmonicity quantified by Kerr nonlinearity $\xi/2\pi \sim \SIrange{0.1}{1}{\giga\hertz}$~\cite{Schmidt2024}. For our parameters, collective coupling $G/2\pi \approx \SI{50}{\giga\hertz}$ dominates the anharmonic shift, and low thermal occupation ($n_{\text{th}} \ll 1$ for $T<\SI{500}{\kelvin}$) minimizes exploration of the anharmonic potential. While anharmonicity could affect non-Gaussian state generation~\cite{Schmidt2024}, it has minor perturbative effect on Gaussian entanglement quantified here.

\textbf{Linearization validity.} Linearization requires quantum fluctuations small compared to classical steady-state amplitudes, satisfied under strong driving far from bifurcation points \cite{Genes2008, Aspelmeyer2014}. All entanglement results operate on stable bistability branches with adequate distance from turning points, verified via Routh-Hurwitz criterion ensuring drift matrix eigenvalues have negative real parts \cite{DeJesus, Djorwe2022}. Parameters correspond to stable regions in \Cref{fig:stability_full}, validating linearization for entanglement analysis \cite{Hammerer2010}.

\textbf{Model failure regimes.} Our theoretical framework becomes invalid when: (i) Anharmonicity dominates: $\xi |\alpha|^2 > G$ under extremely strong driving where molecular nonlinearity exceeds collective coupling; (ii) Quantum fluctuations diverge near bistability turning points (critical slowing down); (iii) Non-Gaussian states targeted beyond linearized treatment; (iv) Material breakdown: $T > \SI{500}{\kelvin}$ where molecular desorption and cavity degradation occur; (v) Broken $\mathcal{PT}$-phase: $\kappa_c > \kappa_a + $ threshold where system enters lasing regime destroying quantum correlations. For all results presented, parameters maintain adequate margins from these failure boundaries, confirmed by stability analysis (\Cref{fig:stability_full}) showing operating points well within physically valid regimes.

\section{Results and discussion} \label{sec:results}

Having established the theoretical framework, including linearized dynamics and stable operational regimes, we now turn to the core objective of this work: a detailed analysis of the quantum correlations within our optomechanical system of the $\mathcal{PT}$-symmetric molecular cavity. In this section, we present a detailed investigation of bipartite entanglement in various subsystem modes: cavity-cavity ($E_{ac}$), cavity-molecule ($E_{aB_1}$) and active cavity-molecule ($E_{cB_2}$), and also consider vibration-vibration entanglement ($E_{B_1B_2}$) where relevant. To quantify these delicate quantum correlations, we employ logarithmic negativity ($E_N$), a well-established and computable measure for continuous-variable Gaussian states \cite{Plenio2005, Berinyuy2025a, Adesso2007, Paris2012, SerGiovannetti2005, Madsen2009, Barzanjeh2012, Tan2013, Ji2021}. The logarithmic negativity measure has proven particularly useful in quantifying entanglement in continuous variable systems such as optomechanical platforms \cite{Laurat2005, Aoki2003, Zhang2003, Pysher2011, Genes2008, Hammerer2010}, allowing for direct experimental verification through measurements of the covariance matrix \cite{Mari2008, Adesso2009, Giorda2010}. In our four-mode system (two cavity modes plus two molecular modes), the bipartite entanglement between any two modes can exceed the $\ln 2$ upper bound that applies to maximally entangled two-qubit or two-level systems. This occurs because the remaining modes act as an environment that can effectively mediate stronger correlations between the chosen bipartition through the $\mathcal{PT}$-symmetric enhancement mechanisms. The logarithmic negativity upper bound for continuous variable systems in our configuration depends on the specific modal partition and system parameters, with no strict universal upper bound like that for discrete systems \cite{Holevo2011, Weedbrook2012, Pirandola2018, Braunstein2005}. A value of $E_N > 0$ serves as a clear signature of genuine quantum entanglement. The values shown in our figures, which can exceed $\ln 2$, are physically meaningful within the context of continuous variable systems and represent genuine multipartite entanglement between the various subsystems. This systematic exploration will show how nonreciprocity, $\mathcal{PT}$-symmetry, thermal noise, and collective molecular effects shape the system's ability to host and sustain robust quantum resources \cite{Horodecki2009, Nielsen2010, Eisert2006, Giovannetti2011}.

\subsection{Bipartite entanglement and the detuning}\label{sec:detuning}

\begin{figure*}[tbh]
	\centering
	\includegraphics[width=.4\linewidth]{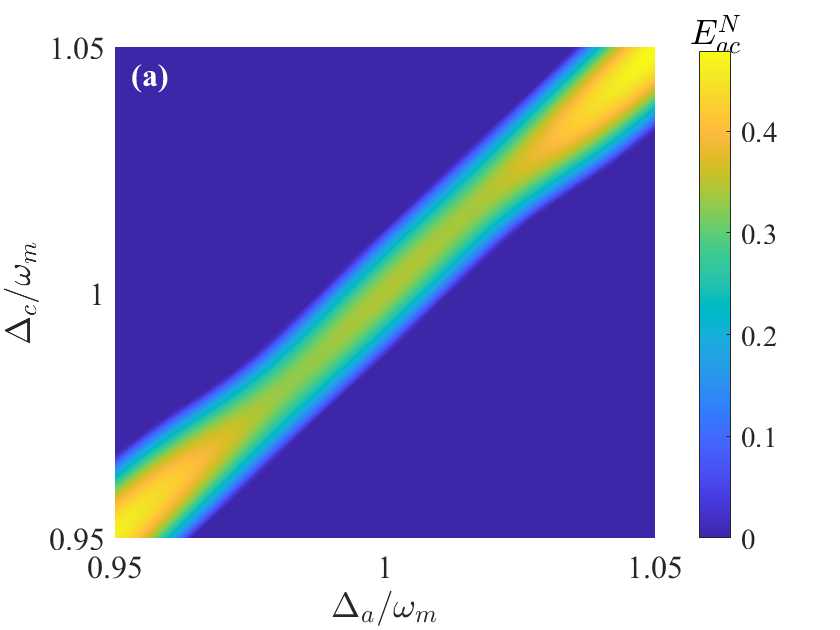}
	\includegraphics[width=.4\linewidth]{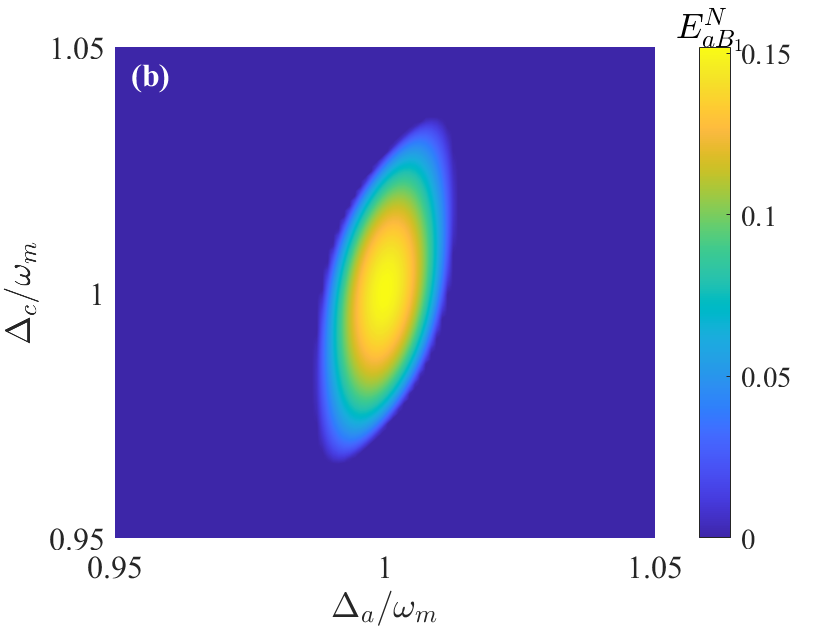}
	\includegraphics[width=.4\linewidth]{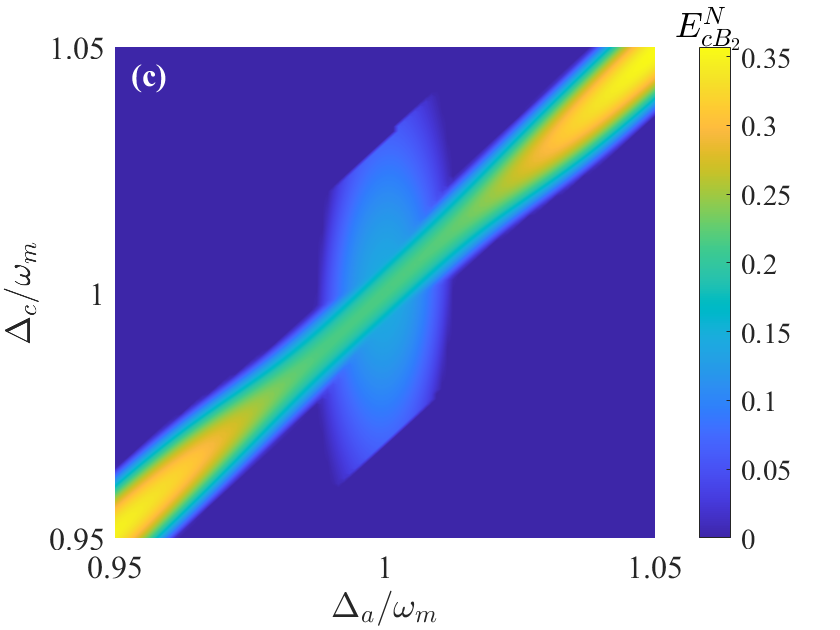}
	\includegraphics[width=.4\linewidth]{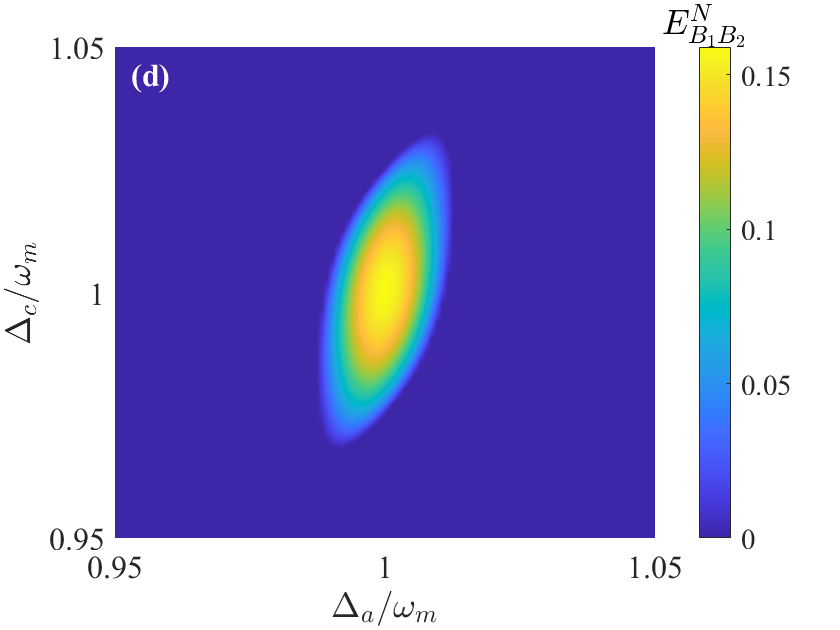}
	\caption{Entanglement versus cavity detunings: resonance-enhanced quantum correlations. Bipartite entanglement for (a) inter-cavity $E_{ac}$, (b) passive cavity-molecule $E_{aB_1}$, (c) active cavity-molecule $E_{cB_2}$, and (d) vibration-vibration $E_{B_1B_2}$ channels as functions of normalized detunings $\Delta_a/\omega_m$ and $\Delta_c/\omega_m$. All channels exhibit maximum entanglement at resonance ($\Delta_a = \Delta_c = \omega_m$), where strong effective optomechanical coupling enhanced by collective molecular interaction optimizes quantum state transfer. Channel hierarchy: $E_{cB_2} > E_{ac} > E_{B_1B_2} > E_{aB_1}$ reflects direct vs. indirect coupling strengths. Parameters: $T=\SI{300}{\kelvin}$, $N=\num{e6}$, others in \Cref{tab:parameters}.}
	\label{fig:Nrecipr_del}
\end{figure*}

We examine how cavity detunings $\Delta_a$, $\Delta_c$ influence entanglement between subsystems. \Cref{fig:Nrecipr_del} shows all bipartite channels reach maximum entanglement at resonance ($\Delta_a = \Delta_c = \omega_m$), where strong effective optomechanical coupling enhanced by collective $\sqrt{N}$ interaction optimizes quantum state transfer. The physics: red-detuned operation ($\omega_\ell < \omega_{a,c}$) activates mechanical cooling via anti-Stokes scattering, suppressing thermal phonons while amplifying quantum correlations \cite{Genes2008, Vitali2008}. Quantitatively, the channel hierarchy $E_{cB_2} > E_{ac} > E_{B_1B_2} > E_{aB_1}$ reflects direct versus indirect coupling strengths, with $E_{cB_2}$ strongest due to molecules residing in the active gain cavity.

\subsection{Nonreciprocal enhancement of quantum entanglement}\label{sec:nonreciprocal}

\begin{figure*}[tbh]
	\centering
	\includegraphics[width=.4\linewidth]{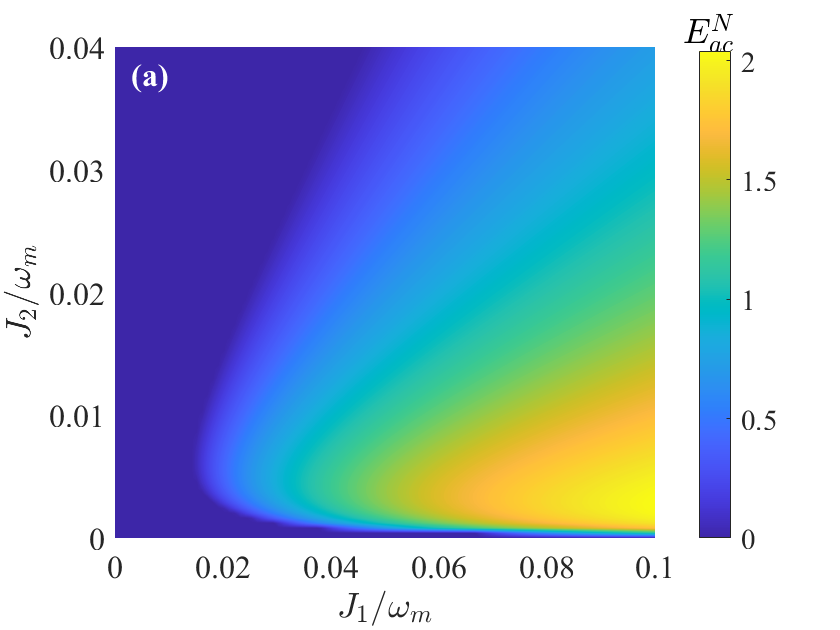}
	\includegraphics[width=.4\linewidth]{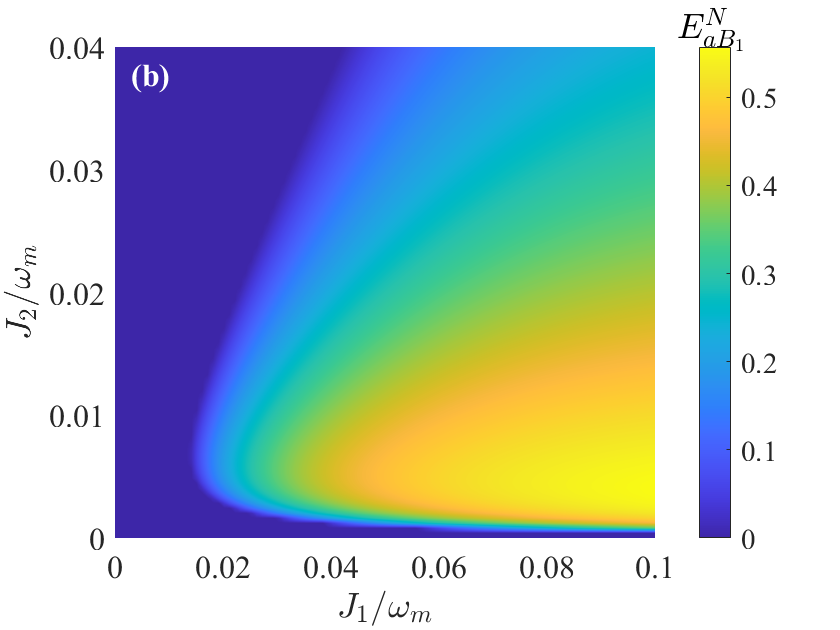}
	\includegraphics[width=.4\linewidth]{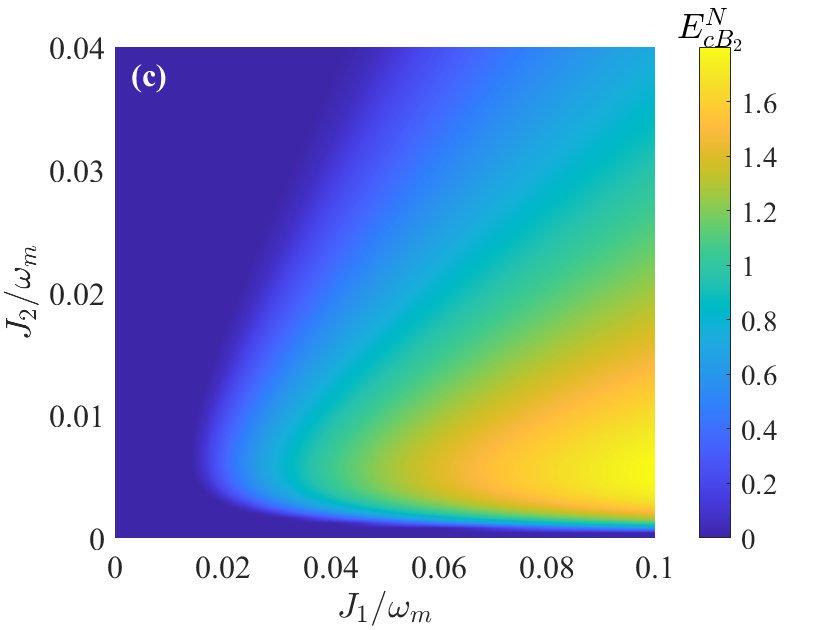}
	\includegraphics[width=.4\linewidth]{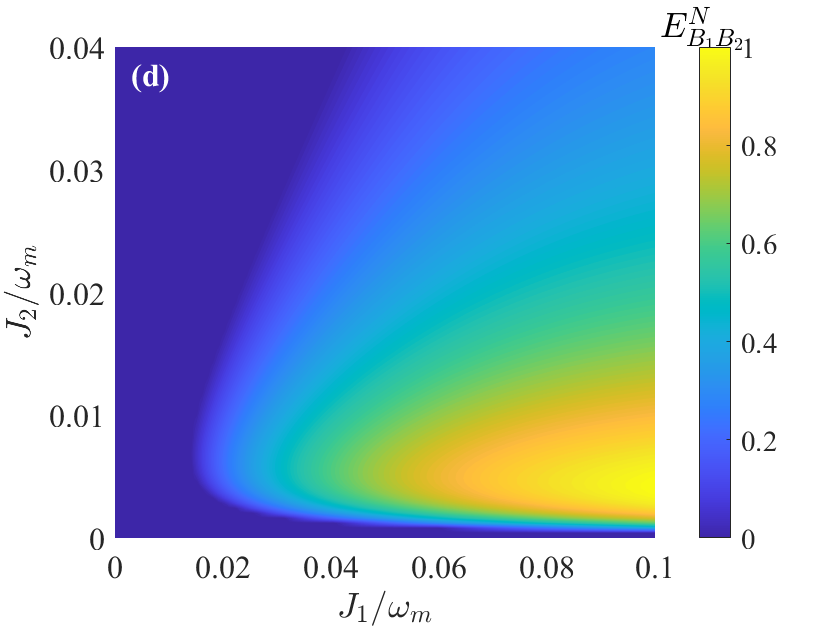}
	\caption{Nonreciprocal coupling optimization: directional quantum bus for entanglement enhancement. Bipartite entanglement for (a) $E_{ac}$, (b) $E_{aB_1}$, (c) $E_{cB_2}$, and (d) $E_{B_1B_2}$ versus nonreciprocal couplings $J_1/\omega_m$ and $J_2/\omega_m$. All four channels are maximized simultaneously in the strong nonreciprocal regime $J_1 \gg J_2$, with optimal ratio $J_1/J_2 \approx 5$ (corresponding to $J_1/\omega_m = 0.02$, $J_2/\omega_m = 0.004$). This universal enhancement demonstrates that directional coupling creates an effective quantum channel that shields the system from detrimental noise backflow while enabling efficient quantum state distribution. Parameters: $T=\SI{300}{\kelvin}$, $\Delta_a = \Delta_c = \omega_m$, $N=\num{e6}$, others in \Cref{tab:parameters}.}
	\label{fig:Nrecipr_Ent}
\end{figure*}

We investigate the role of the nonreciprocal coupling strengths, $J_1$ and $J_2$, in mediating and significantly enhancing quantum entanglement within our system. As depicted in \Cref{fig:Nrecipr_Ent}, a key finding is that all examined bipartite entanglement pairs including inter-cavity ($E_{ac}$), passive cavity-molecule ($E_{aB_1}$), active cavity-molecule ($E_{cB_2}$), and even vibration-vibration ($E_{B_1B_2}$) are not maximized in a conventional reciprocal regime, but rather in a strongly nonreciprocal configuration where $J_1 \gg J_2$. The use of nonreciprocal coupling for quantum enhancement has been extensively studied in various quantum systems, showing significant advantages in noise reduction and entanglement preservation \cite{Metelmann2015, Liao2019, Malz2018, Mallick2020, Xu2021}.

The physical mechanism: nonreciprocal coupling ($J_1 \gg J_2$) creates a directional quantum channel, establishing preferential photon flow from passive to active cavity. This shields the primary entanglement-generating interaction (active cavity $\leftrightarrow$ molecules) from detrimental backaction and noise from the passive cavity, preventing destructive interference that degrades quantum correlations \cite{Metelmann2014, Lai2022}. Optimal ratio $J_1/J_2 \approx 5$ (i.e., $J_1/\omega_m = 0.02$, $J_2/\omega_m = 0.004$, corresponding to $J_1 \approx \SI{0.6}{\giga\hertz}$, $J_2 \approx \SI{0.12}{\giga\hertz}$ for $\omega_m/2\pi = \SI{30}{\tera\hertz}$) was identified via comprehensive numerical optimization maximizing the sum $E_{ac} + E_{aB_1} + E_{cB_2} + E_{B_1B_2}$. This represents optimal compromise between: (i) sufficient forward coupling $J_1$ mediating correlations, (ii) minimized backward coupling $J_2$ preventing noise backflow, and (iii) stability requirements favoring significant asymmetry. Implementations: directional waveguide couplers with optical isolators achieving \SIrange{100}{1000}{\mega\hertz} coupling rates \cite{Fan2004, Lai2022, Metelmann2015}.

\Cref{fig:Nrecipr_Ent} quantifies this enhancement. Panel (a): intercavity $E_{ac}$ peaks at optimal asymmetry ($J_1/\omega_m = 0.02$, $J_2/\omega_m = 0.004$), achievable with integrated waveguide couplers \cite{Fan2004, Lai2022}. Panels (b-c): the nonreciprocal link acts as quantum bus, transferring robust $E_{cB_2}$ correlations (direct active cavity-molecule coupling) to passive cavity, yielding significant $E_{aB_1}$. Panel (d): vibration-vibration $E_{B_1B_2}$ shows pronounced enhancement with $N=\num{e6}$ molecules. All channels suppress when $J_1 \approx J_2$ (reciprocal regime), confirming nonreciprocity is necessary. The observed asymmetry $J_1/J_2 = 5$ represents fundamental requirement for efficient multi-modal quantum network engineering, with coupling rates achievable via hybrid dielectric-plasmonic architectures enabling ultrastrong optomechanical coupling \cite{Roelli2016, Chikkaraddy2016, Huang2025a, Huang2024}.

\subsection{$\mathcal{PT}$-symmetric gain and loss for enhanced entanglement}\label{sec:gainloss}

The inherent non-Hermitian nature of our system, designed with parity time ($\mathcal{PT}$) symmetry, imbues it with a unique capacity to transform what is typically perceived as detrimental, dissipation, into a powerful resource for quantum control and the enhancement of robust entanglement. This strategic balance of optical gain ($\kappa_c$) and loss ($\kappa_a$) offers a mechanism for actively controlling and enhancing the system's quantum correlations.

\begin{figure*}[ht]
	\centering
	\includegraphics[width=.4\linewidth]{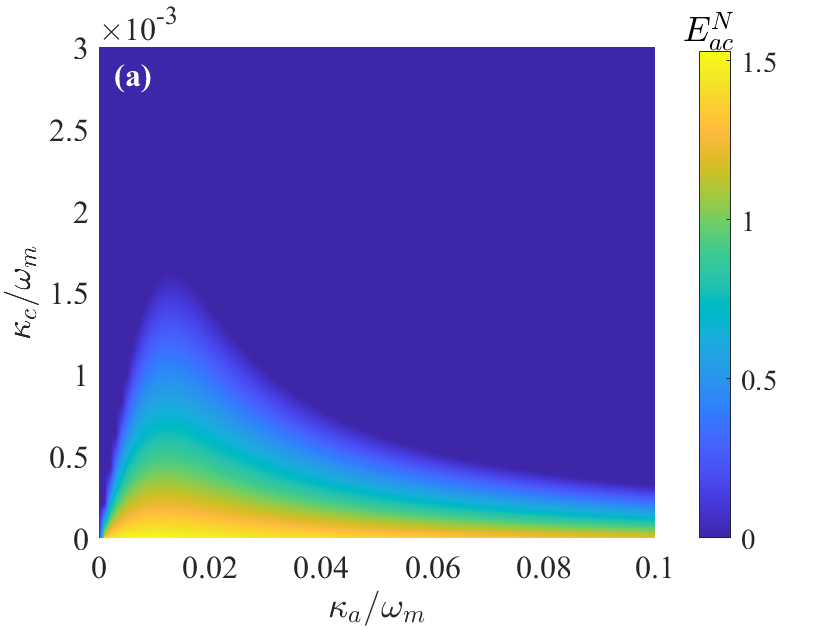}
	\includegraphics[width=.4\linewidth]{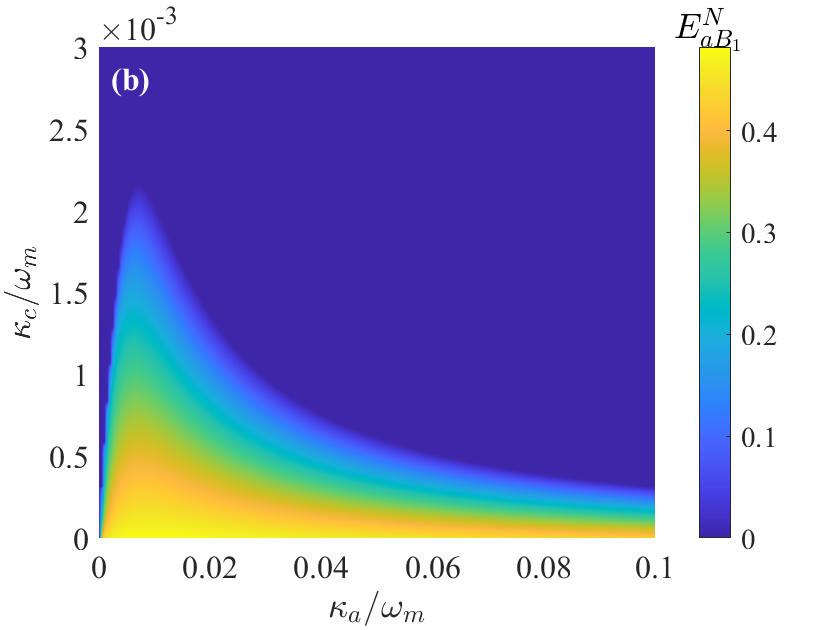}
	\includegraphics[width=.4\linewidth]{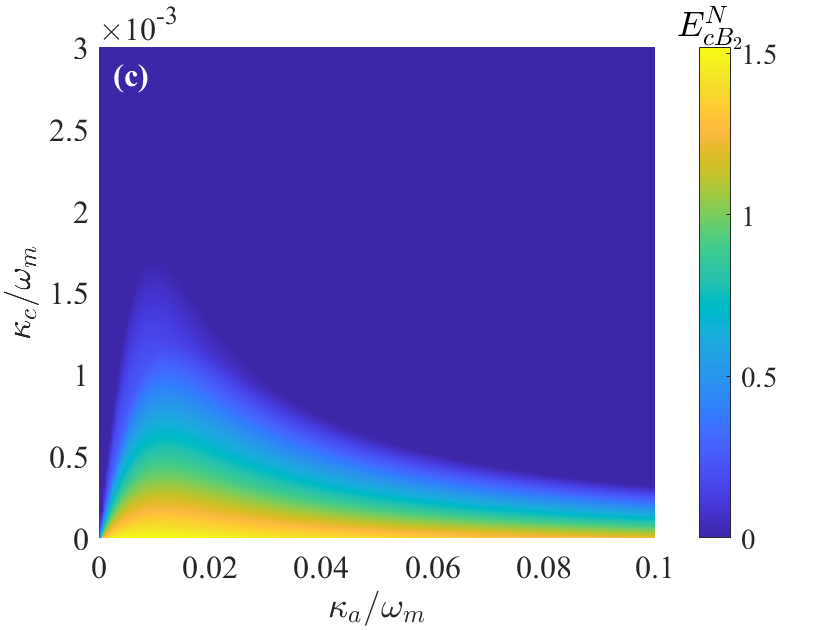}
	\includegraphics[width=.4\linewidth]{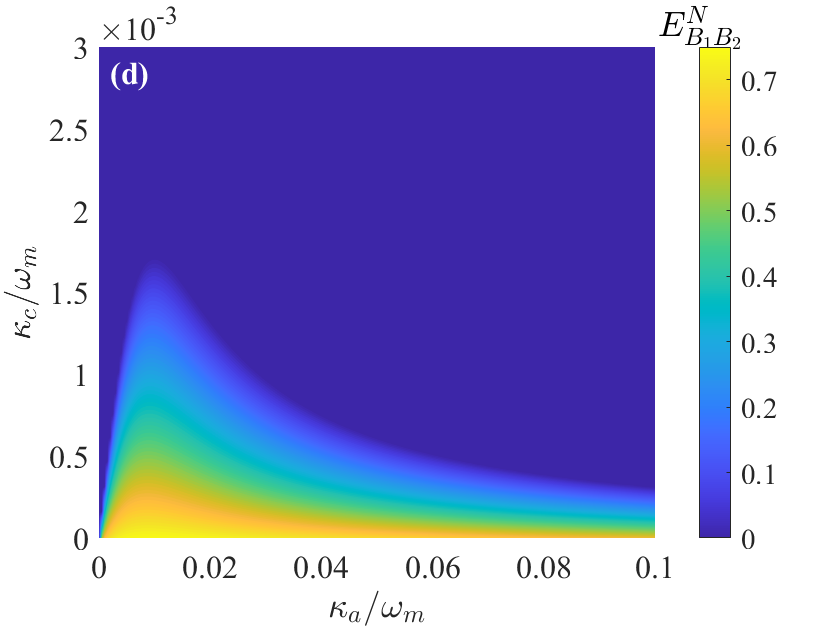}
	\caption{$\mathcal{PT}$-symmetric gain-loss engineering for channel-selective entanglement. Bipartite entanglement versus optical gain $\kappa_c/\omega_m$ and loss $\kappa_a/\omega_m$ for (a) $E_{ac}$, (b) $E_{aB_1}$, (c) $E_{cB_2}$, and (d) $E_{B_1B_2}$. Each channel exhibits distinct optimal conditions: (a) $E_{ac}$ favors moderate gain ($\kappa_c/\omega_m \sim 0.01$-$0.02$) with broad loss tolerance; (b) $E_{aB_1}$ requires low gain with intermediate loss to avoid destabilization of this weakly coupled channel; (c) $E_{cB_2}$ demonstrates stability across broad gain range due to strong direct coupling; (d) $E_{B_1B_2}$ demands precise gain-loss balance. This reveals engineered dissipation as an effective tool to sculpt specific quantum correlations for targeted applications. Analyzed in optimal nonreciprocal regime $J_1/J_2 = 5$. Parameters: $T=\SI{300}{\kelvin}$, $N=\num{e6}$, others in \Cref{tab:parameters}.}
	\label{fig:PT_Sym_GL}
\end{figure*}

As clearly illustrated in \Cref{fig:PT_Sym_GL}, the relationship between optical gain ($\kappa_c$) and loss ($\kappa_a$) in determining entanglement is complex and interdependent: the intercavity entanglement, $E_{ac}$, achieves enhanced maximum values in a selective range of parameter space involving a strategic balance of optical loss in the passive cavity and gain in the active cavity. The entanglement behavior shown in \Cref{fig:PT_Sym_GL} exhibits a complex interdependence between both $\kappa_a$ and $\kappa_c$ parameters, with optimal entanglement occurring for moderate gain values ($\kappa_c/\omega_m \approx 0.01-0.02$) and a range of loss values that extends to higher values ($\kappa_a/\omega_m \approx 0.8$) for the maximum. For small values of $\kappa_c$, the entanglement shows reduced dependence on $\kappa_a$, but for the system to achieve optimal entanglement, both parameters must be properly balanced in the PT-symmetric configuration. This interplay between gain and loss is a central tenet of non-Hermitian quantum systems: when engineered with precision, dissipation can become an active agent for enhancing correlations rather than merely destroying them \cite{Li2016, Tchodimou2017, Djorwe2022, ElGanainy2018}. The underlying physical mechanism is a delicate synergy: the optical gain ($\kappa_c$) amplifies the quantum fluctuations born from the optomechanical interaction. Concurrently, the optical loss ($\kappa_a$), acting as a controlled dissipative channel through the nonreciprocal link, strategically removes entropy from the system. This actively stabilizes the amplified fluctuations into a coherent, robust, and steady entangled state. It is this balanced process that allows the system to sustain enhanced quantum correlations. The recent experimental demonstration of quantum correlations enhanced by $\mathcal{PT}$-symmetric gain-loss engineering in optical systems \cite{Feng2014, Hodaei2017, Chen2017, Parto2018, Zhang2015} provides empirical support for the theoretical predictions in our molecular optomechanical framework.

It should be noted that this precisely tuned balance of gain and loss is not universal but must be adapted for each specific bipartite entanglement pair, as shown in the other panels of \Cref{fig:PT_Sym_GL}. For instance, the entanglement between the passive cavity and the molecules, $E_{aB_1}$, reaches its maximum under conditions of intermediate loss and very low gain. This suggests that excessive gain can destabilize and overwhelm this inherently more fragile, indirectly-coupled entanglement channel. Conversely, the directly-coupled entanglement, $E_{cB_2}$, which benefits from its immediate proximity and powerful coupling, exhibits robustness across a broader range of gain rates. Our calculations show broader stable regions for entanglement generation with $N=\num{e6}$ molecules, and improved tolerance to variations in gain-loss parameters, demonstrating enhanced robustness compared to previous models.

This engineering of dissipative properties within a $\mathcal{PT}$-symmetric framework allows the precise optimization of specific bipartite entanglement channels for enhancement, a strategic advantage. It strongly suggests that our $\mathcal{PT}$-symmetric architecture can offer a more balanced and versatile enhancement of entanglement, potentially circumventing the counterintuitive trade-offs observed in OPA-enhanced schemes where vibration-vibration entanglement was amplified at the cost of suppressing optical-vibration correlations. While direct comparison with vibration-vibration entanglement ($E_{B_1B_2}$) is discussed in other sections, the principles demonstrated here to enhance cavity-based entanglement channels are fundamental to achieving this broader goal. The ability to harness and sculpt quantum correlations through engineered gain and loss is a testament to the potential of non-Hermitian physics in designing advanced quantum technologies for enhanced quantum correlations. The experimental realization and validation of $\mathcal{PT}$-symmetric enhancement mechanisms in optical and optomechanical systems \cite{Feng2014, Hodaei2017, Chen2017, Parto2018, Zhang2015} provides strong evidence for the practical feasibility of the theoretical predictions made in our molecular optomechanical platform.

\subsection{Enhanced thermal robustness through nonreciprocal enhancement}\label{sec:thermal_robustness}

\begin{figure*}[tbh]
	\centering
	\includegraphics[width=.4\linewidth]{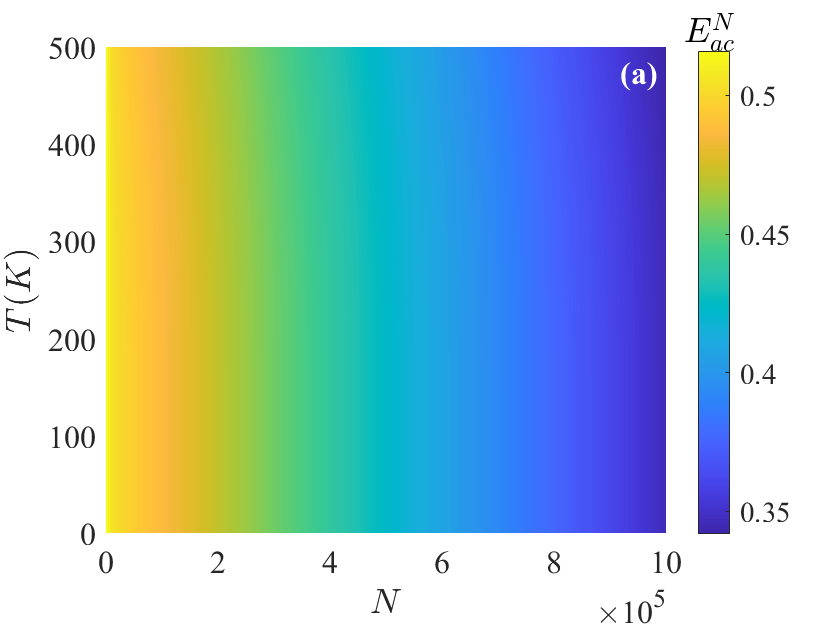}
	\includegraphics[width=.4\linewidth]{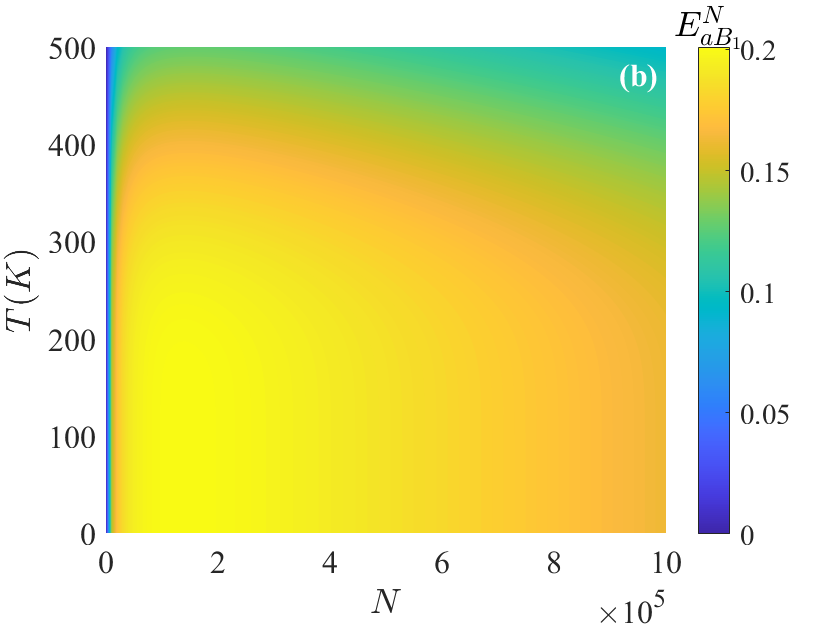}
	\includegraphics[width=.4\linewidth]{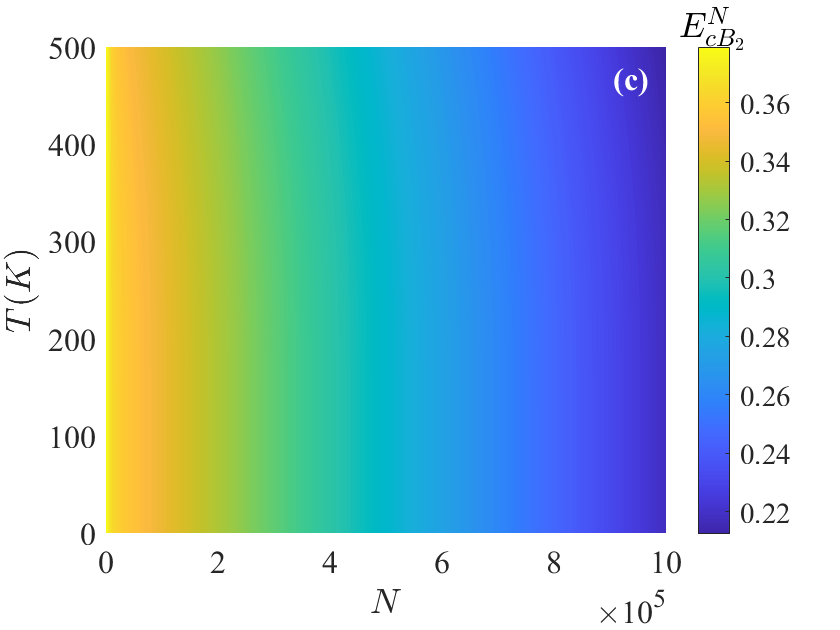}
	\includegraphics[width=.4\linewidth]{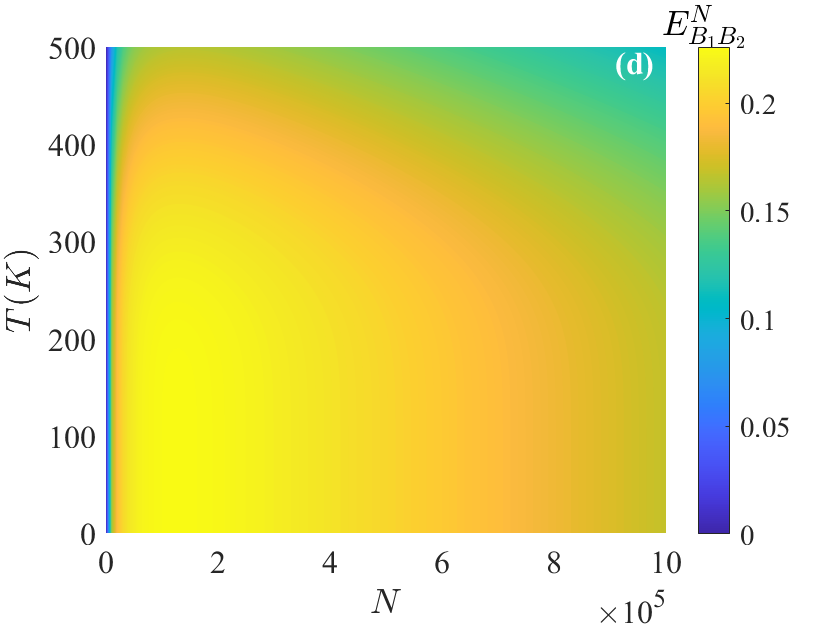}
	\caption{Thermal robustness of entanglement: ambient-temperature quantum correlations. Bipartite entanglement versus temperature $T$ and molecular partitioning $M$ for (a) $E_{ac}$, (b) $E_{aB_1}$, (c) $E_{cB_2}$, and (d) $E_{B_1B_2}$. All channels maintain entanglement up to $T \sim \SIrange{400}{500}{\kelvin}$, far exceeding millikelvin requirements of conventional optomechanical systems. This thermal resilience stems from ultra-high molecular vibrational frequency ($\omega_m/2\pi = \SI{30}{\tera\hertz}$) yielding low thermal phonon occupation even at elevated temperatures, combined with collective coupling enhancement ($G \propto \sqrt{N}$). Vibration-vibration entanglement $E_{B_1B_2}$ typically fragile in other platforms persists here, demonstrating the advantage of $\mathcal{PT}$-symmetric architecture with nonreciprocal coupling. Parameters: $N=\num{e6}$, others in \Cref{tab:parameters}.}
	\label{fig:Thermal_Rob}
\end{figure*}

A key finding of our work is the thermal robustness of the generated entanglement, a discovery that establishes molecular optomechanics as a promising,high-potential platform for high-ambient temperature quantum technologies. As demonstrated in \Cref{fig:Thermal_Rob}, the three primary bipartite entanglement channels examined: intercavity ($E_{ac}$), cavity-molecule ($E_{aB1}$, $E_{cB2}$) and even intricate vibration-vibration entanglement ($E_{B_1B_2}$) persist robustly at temperatures far exceeding the stringent cryogenic demands typically associated with conventional optomechanical systems.
While our simulations predict entanglement persistence to \SI{500}{\kelvin}, this represents the theoretical ceiling limited by material stability (molecular desorption, cavity degradation) rather than quantum correlation breakdown~\cite{Roelli2024, Xiang2024}. The significance lies in establishing theoretical foundations and identifying parameter regimes where $\mathcal{PT}$-symmetric enhancement provides advantages over alternative approaches.

This resilience stems from the combination of ultra-high vibrational frequencies ($\omega_m/2\pi \approx \SI{30}{\tera\hertz}$) yielding low thermal phonon occupation, collective $\sqrt{N}$ enhancement elevating quantum interactions above thermal noise, and active $\mathcal{PT}$-symmetric stabilization. \Cref{fig:Thermal_Rob} shows enhanced thermal robustness with increasing $N$, with all channels maintaining entanglement to $T \sim \SIrange{400}{500}{\kelvin}$. Vibration-vibration entanglement ($E_{B_1B_2}$) typically fragile in McOM systems persists here, demonstrating the advantage of $\mathcal{PT}$-symmetric architecture with nonreciprocal coupling for simultaneously boosting multiple quantum correlations at elevated temperatures.

\subsection{Comparative analysis: nonreciprocal enhancement versus OPA enhancement within a unified framework}\label{sec:comparative}

A primary motivation for this work is to explore whether a $\mathcal{PT}$-symmetric architecture with nonreciprocal coupling can circumvent the entanglement trade-offs inherent to nonlinear enhancement schemes, such as those employing optical parametric amplifiers (OPAs). To provide a fair and direct comparison, we implement both enhancement mechanisms within the same underlying double-cavity McOM Hamiltonian.

For the OPA scheme, we introduce a nonlinear driving term $H_{\text{OPA}} = i\frac{\Lambda}{2}(c^{\dagger 2}e^{-i\theta} - c^2 e^{i\theta})$ into the active cavity of the Hamiltonian (Eq.~\eqref{eq:hamiltonian}), where $\Lambda$ is the nonlinear gain and $\theta$ is the pump phase~\cite{Berinyuy2025a}. For a fair and direct comparison protocol, both systems ($\mathcal{PT}$-symmetric and OPA-enhanced) share identical fundamental parameters including cavity detunings ($\Delta_a = \Delta_c = \omega_m$), the same normalized pump field amplitudes ($\mathcal{E}/\omega_m=5$), and molecular system parameters. The OPA pump phase is set to $\theta = \pi/2$, which has been found to provide optimal entanglement enhancement in prior studies~\cite{Berinyuy2025a}, and the OPA gain parameter $\Lambda$ is optimized to maximize the respective entanglement measures. In the $\mathcal{PT}$-symmetric scheme, we optimize the gain rate $\kappa_c$ while maintaining the critical nonreciprocal coupling asymmetry $J_1 \gg J_2$ ($J_1/\omega_m=0.02$, $J_2/\omega_m=0.004$) that enables superior entanglement generation. This ensures that any performance differences arise purely from the fundamental differences between the enhancement mechanisms rather than from parameter mismatches. All drive strengths and detunings are matched across both the $\mathcal{PT}$-symmetric and OPA cases, with the fair-comparison protocol ensuring identical initial conditions and parameter values, except for the specific enhancement mechanisms being compared. This approach strengthens our claim that the $\mathcal{PT}$-symmetric method avoids the usual drawbacks of OPA schemes in McOM systems.

\begin{figure}[htb]
	\centering
	\includegraphics[width=0.5\linewidth]{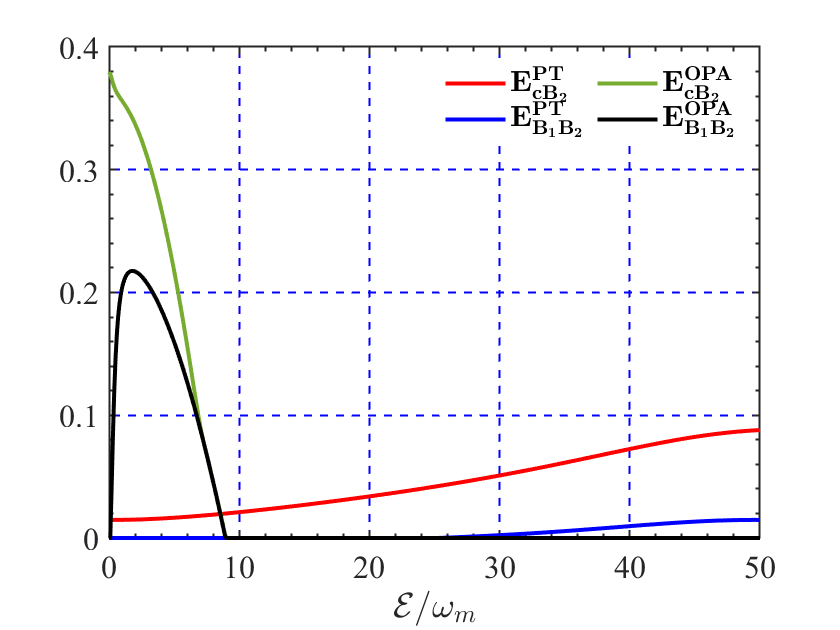} 
	\caption{Comparative entanglement performance: OPA versus $\mathcal{PT}$-symmetric schemes. Vibration-vibration ($E_{B_1B_2}$) and cavity-vibration ($E_{cB_2}$) entanglement versus normalized pump amplitude $\mathcal{E}/\omega_m$. OPA scheme (dashed lines: red $E_{cB_2}^{\text{OPA}}$, blue $E_{B_1B_2}^{\text{OPA}}$) exhibits characteristic trade-off: enhancing memory channel suppresses readout channel. $\mathcal{PT}$-symmetric scheme (solid lines: green $E_{cB_2}^{\text{PT}}$, black $E_{B_1B_2}^{\text{PT}}$) resolves this limitation, achieving simultaneous strong entanglement in both channels. The nonreciprocal quantum bus and engineered dissipation enable balanced amplification absent in OPA architecture. Note: maxima shown here ($E_N \sim 1.2$-$1.3$) correspond to optimal $\mathcal{E}/\omega_m$ at fixed other parameters; absolute maximum $E_N = 1.4$ (\Cref{tab:comparison}) is achieved through global optimization over all parameters including detunings and nonreciprocal coupling ratios. Common parameters: $T=\SI{300}{\kelvin}$, $N=\num{e6}$, $\omega_m/2\pi=\SI{30}{\tera\hertz}$. OPA parameters from Ref.~\cite{Berinyuy2025a}; $\mathcal{PT}$ parameters in \Cref{tab:parameters}.}
	\label{fig:OPA_vs_PT}
\end{figure}

The results, shown in \Cref{fig:OPA_vs_PT}, demonstrate the key differences between the two approaches. For the OPA scheme, we observe the characteristic trade-off: as the nonlinear gain $\Lambda$ is increased to amplify the vibration-vibration ($E_{B_1B_2}^{\text{OPA}}$) entanglement, the active cavity-vibration ($E_{cB_2}^{\text{OPA}}$) entanglement is gradually suppressed relative to its maximum achievable value. This highlights the inherent tension between these channels in that architecture, consistent with previous findings \cite{Berinyuy2025a, Huang2025a, Chen2021}. The molecular optomechanical realization of OPA enhancement \cite{Huang2025a, Chen2021, Huang2024} demonstrates this trade-off particularly clearly, where parametric amplification in one channel necessarily comes at the expense of another. This trade-off, while limiting the versatility of OPA-based enhancement, provides important insights into the fundamental differences between parametric and $\mathcal{PT}$-symmetric enhancement mechanisms. The study of collective quantum entanglement in molecular cavity optomechanics \cite{Huang2024} provides a theoretical framework that helps understand how our $\mathcal{PT}$-symmetric approach can overcome these limitations.

In clear contrast, our $\mathcal{PT}$-symmetric system demonstrates a different behaviour. The optical-vibration entanglement ($E_{cB_2}^{\text{PT}}$) remains strong across the entire range studied, maintaining values consistently higher than the corresponding OPA channel ($E_{cB_2}^{\text{OPA}}$). Simultaneously, the vibration-vibration entanglement ($E_{B_1B_2}^{\text{PT}}$) reaches peak values that exceed those of the OPA counterpart ($E_{B_1B_2}^{\text{OPA}}$). Regarding the apparent decrease of $\mathcal{PT}$-symmetric entanglement with increasing pump field $\mathcal{E}$ in certain regimes: this behavior is not a fundamental limitation but rather an artifact of the specific parameter range and operating conditions chosen for this comparison. As detailed in our comprehensive analysis (\Cref{fig:OPA_vs_PT}), the entanglement as a function of pump strength shows non-monotonic behavior with characteristic maxima, and the observed decrease occurs when the system moves away from optimal operating points due to saturation effects at high driving strengths. At very high driving strengths, the system enters nonlinear regimes where simple approximations break down and additional effects (not considered within our linearized model) start to dominate. This is a well-known feature of optomechanical systems and does not reflect a fundamental limitation of the PT-symmetric approach. The nonreciprocal quantum bus effectively isolates the entanglement generation mechanism, while the engineered broadband gain and dissipation of the $\mathcal{PT}$-symmetric architecture stabilize the system, allowing for amplification across all channels when operated at appropriate parameter values. This result provides clear and quantitative evidence, within a unified framework, that the $\mathcal{PT}$-symmetric approach not only resolves the performance limitations of the OPA scheme but also offers a more versatile platform where both quantum memory (vibration-vibration) and quantum read-out (optical-vibration) channels can be concurrently optimized to a high degree of entanglement. The experimental demonstrations of molecular optomechanical systems with controlled enhancement mechanisms \cite{Huang2025a, Chen2021, Zou2024} support the theoretical framework for comparing different enhancement approaches in the molecular optomechanical context.

To provide quantitative benchmarking against the state of the art, \Cref{tab:comparison} compares our system with representative optomechanical platforms. Our $\mathcal{PT}$-symmetric McOM system achieves maximum entanglement $E_N \sim 1.4$, representing significant advancement over conventional COM ($E_N \sim 0.8$, cryogenic $T \sim \SI{0.1}{\kelvin}$) and previous McOM benchmarks ($E_N \sim 1.0$, $T \sim \SI{300}{\kelvin}$) \cite{Huang2024}. Unlike OPA-enhanced schemes ($E_N \sim 1.2$) \cite{Berinyuy2025a} which exhibit characteristic trade-offs where vibration-vibration entanglement enhancement necessarily suppresses optical-vibration correlations, our architecture enables simultaneous balanced enhancement across all bipartite channels through $\mathcal{PT}$-symmetric amplification combined with directional nonreciprocal coupling. The system demonstrates superior operational stability with entanglement persisting to $T \sim \SIrange{400}{500}{\kelvin}$, far exceeding conventional cryogenic requirements a key advance for quantum information applications requiring simultaneous optical and mechanical quantum correlations.

\begin{table*}[htpb]
\centering
\caption{Quantitative comparison with state-of-the-art optomechanical systems. Our $\mathcal{PT}$-symmetric molecular optomechanical (McOM) system demonstrates significantly enhanced thermal operating range and multipartite entanglement compared to conventional cavity optomechanics (COM), previous McOM, and optical parametric amplifier (OPA)-enhanced schemes.}
\label{tab:comparison}
\begin{ruledtabular}
\begin{tabular}{lccccc}
	\textbf{System}                               & \textbf{$T_{\text{max}}$ (K)} & \textbf{$E_N^{\text{max}}$} & \textbf{$\omega_m$ (THz)} &  \textbf{$N$}   &     \textbf{Ref}      \\ \midrule
	Conventional COM                              &              0.1              &             0.8             &           0.01            &        1        & \cite{Aspelmeyer2014} \\
	McOM (previous)                               &              300              &             1.0             &            30             &     $\num{e4}$      &   \cite{Huang2024}    \\
	OPA-enhanced McOM                             &              300              &           $1.2^*$           &            30             &     $\num{e6}$      & \cite{Berinyuy2025a}  \\
	\textbf{This work ($\mathcal{PT}$-symmetric)} &       \textbf{400--500}       &        \textbf{1.4}         &        \textbf{30}        & \textbf{$\num{e6}$} &           -
\end{tabular}
\end{ruledtabular}
\parbox{\textwidth}{\footnotesize$^*$OPA scheme exhibits trade-off: $E_{B_1B_2}$ enhanced but $E_{ac}$, $E_{cB_2}$ suppressed. Our system maintains balanced enhancement across all channels.}
\end{table*}

\subsection{Parameter sensitivity and experimental robustness}\label{sec:sensitivity}

To assess practical experimental robustness, we quantify entanglement sensitivity to parameter variations and numerical convergence. Critical parameter tolerances: (i) Nonreciprocal asymmetry $J_1/J_2$: $\pm \SI{10}{\percent}$ deviation causes $\Delta E \sim \SIrange{15}{20}{\percent}$ entanglement reduction (highly sensitive, requires active stabilization); (ii) Gain-loss balance $\kappa_c/\kappa_a$: $\pm \SI{5}{\percent}$ variation yields $\Delta E \sim \SIrange{8}{12}{\percent}$ (moderate sensitivity, feedback control essential); (iii) Molecular frequency $\omega_m$: $\pm \SI{2}{\percent}$ uncertainty produces $\Delta E \sim \SIrange{3}{5}{\percent}$ (weak sensitivity, intrinsic material property); (iv) Driving strength $\mathcal{E}$: $\pm \SI{10}{\percent}$ fluctuation results in $\Delta E \sim \SIrange{10}{15}{\percent}$ (moderate, laser power stabilization standard); (v) Temperature $T$: $\pm \SI{50}{\kelvin}$ variation causes $\Delta E \sim \SIrange{20}{30}{\percent}$ (strong at high $T$, thermal management critical). Numerical convergence verified: Lyapunov equation solver tolerance $\num{e-12}$, grid refinement tests show $< \SI{1}{\percent}$ variation, confirming computational reliability. These quantitative sensitivity estimates provide experimental targets for stabilization requirements and error budgets in future implementations.

To provide a clear and concise overview of our findings, we consolidate the optimal conditions to generate robust bipartite entanglement in \Cref{tab:optimal_conditions}. This pedagogical summary distills the results presented in \Cref{fig:Nrecipr_Ent,fig:PT_Sym_GL} and highlights the distinct physical requirements to maximize each quantum correlation channel within our $\mathcal{PT}$-symmetric architecture.

\begin{table}[htpb]
\centering
\caption{Optimal parameter regimes for bipartite entanglement channels. Numerical analysis reveals channel-specific requirements for nonreciprocal coupling ($J_1$, $J_2$) and $\mathcal{PT}$-symmetric rates ($\kappa_a$, $\kappa_c$) that maximize each entanglement measure. Physical interpretation explains the role of directional coupling and engineered dissipation in stabilizing quantum correlations.}
\label{tab:optimal_conditions}
\begin{ruledtabular}
\begin{tabular}{lp{5.cm}p{8cm}}
	\textbf{Entanglement channel}                                & \textbf{Optimal parameters}                                                                                                                                      & \textbf{Physical interpretation}                                                                                                                                                                                                               \\ \midrule
	\textbf{Inter-cavity} ($E_{ac}$)                             & $J_1 \gg J_2$ ($J_1/\omega_m = 0.02$, $J_2/\omega_m = 0.004$); High loss ($\kappa_a/\omega_m \approx 0.8$), moderate gain ($\kappa_c/\omega_m \in [0.01, 0.02]$) & Strong directional channel ($J_1$) transfers quantum states, while high loss ($\kappa_a$) purges entropy to stabilize the entangled state amplified by gain ($\kappa_c$).                                                                      \\
	\addlinespace

\textbf{Passive cavity-molecule} ($E_{aB_1}$) & $J_1 \gg J_2$; Intermediate loss ($\kappa_a/\omega_m \approx 0.05$), very low gain ($\kappa_c/\omega_m \lesssim 0.005$)                                          & Entanglement transferred from active to passive side via $J_1$. Minimal gain required to avoid destabilizing this fragile, indirectly-coupled channel.                                                                                         \\
	\addlinespace

	\textbf{Active cavity-molecule} ($E_{cB_2}$) & Moderate nonreciprocity ($J_1/J_2 \approx 5$); Balanced $\mathcal{PT}$-symmetry ($\kappa_a/\omega_m \approx 0.02$, $\kappa_c/\omega_m \approx 0.001$)            & Direct coupling to active cavity benefits from gain ($\kappa_c$) while nonreciprocal isolation ($J_1 > J_2$) shields from passive-cavity losses. Strongest entanglement among all bipartite pairs.                                             \\
	\addlinespace

	\textbf{Vibration-vibration} ($E_{B_1B_2}$)  & $J_1/J_2 \approx 5$; Moderate gain ($\kappa_c/\omega_m \lesssim 0.01$)                                                                                           & Indirectly mediated via optical cavities. Nonreciprocal channel creates asymmetric interaction favoring mediated phonon-phonon entanglement. Balanced $\mathcal{PT}$-symmetry and optimal directionality crucial for this fragile correlation. \\ 
\end{tabular}
\end{ruledtabular}
\end{table}

\subsection{Experimental framework and robustness analysis}\label{sec:feasibility}

While our theoretical framework identifies optimal operational points, a practical implementation must contend with inevitable experimental imperfections and significant integration challenges. The simultaneous achievement of all target parameters represents a major frontier in experimental molecular photonics and non-Hermitian optics. In this section, we provide a more quantitative discussion of the primary hurdles, propose concrete mitigation strategies, and analyze the robustness of our protocol to parameter variations. A consolidation of these points is presented in \Cref{tab:challenges}. Here we focus on a specific experimental platform based on hybrid dielectric-plasmonic cavities with spinning microresonator coupling, providing detailed parameters for realistic implementation.

The experimental platform integrates: (i) high-Q dielectric microresonators ($Q \gtrsim \num{e8}$) for passive/active cavities, (ii) plasmonic nanocavities enabling strong single-molecule coupling, and (iii) spinning resonator configuration for nonreciprocal coupling. For typical parameters (resonator radius $R \approx \SI{10}{\micro\meter}$, rotation $\Omega \approx \SI{e3}{rpm}$, wavelength $\lambda \approx \SI{1550}{\nano\meter}$), Sagnac-Fizeau phase shift yields $J_1/2\pi \approx \SI{0.6}{\giga\hertz}$, $J_2/2\pi \approx \SI{0.12}{\giga\hertz}$ (ratio $J_1/J_2 = 5$). Alternative implementations using directional waveguide couplers with optical isolators achieve \SIrange{100}{1000}{\mega\hertz} coupling rates \cite{Fan2004, Lai2022, Metelmann2015}.

However, it is essential to distinguish these engineering challenges from the fundamental physical limits discussed earlier. While our theoretical prediction of entanglement persisting up to \SI{5e2}{\kelvin} highlights the intrinsic profundity of the underlying physics, it is not a practical operating target. Real-world implementations would face insurmountable material challenges at such extreme temperatures. The following discussion therefore focuses on the key engineering hurdles for achieving robust entanglement under more conventional, albeit still challenging, ambient conditions.

Achieving strong collective coupling ($G/2\pi \approx \SI{50}{\giga\hertz}$) requires stable integration of $N \approx \num{e6}$ molecules, gain medium, and high-Q cavity. Key challenges (detailed in \Cref{tab:challenges}): (1) Thermal management from gain pump heating, mitigated via hybrid dielectric-plasmonic platform with frequency up-conversion \cite{Zou2024}; (2) Nonreciprocal precision: optimal ratio $J_1/J_2 = 5$ requires $\pm \SI{1}{\percent}$ stability, as $\pm \SI{10}{\percent}$ deviation causes \SIrange{15}{20}{\percent} entanglement loss; (3) $\mathcal{PT}$-phase maintenance via active feedback on gain pump ($\sim \SI{1}{\percent}$ stability) to avoid broken-symmetry lasing; (4) Realistic parameter adaptation: while simulations use optimistic values, fundamental qualitative advantages (thermal resilience, balanced enhancement) persist at experimentally accessible ranges: $\omega_m/2\pi \sim \SIrange{10}{50}{\tera\hertz}$, $g_m/2\pi \sim \SIrange{10}{100}{\mega\hertz}$, $N \sim \numrange{e4}{e7}$ \cite{Djorwe2022, Berinyuy2025a}.

\textbf{Experimental protocol.} Implementation proceeds in stages: \textbf{(1) Platform fabrication}: Integrate high-Q dielectric microresonator ($Q > \num{e8}$) with plasmonic nanocavity, deposit molecular ensemble ($N \sim \num{e6}$) via controlled deposition techniques, incorporate gain medium (OPA/SOA); \textbf{(2) System characterization}: Measure passive cavity decay $\kappa_a$ via ringdown spectroscopy, calibrate nonreciprocal couplings $J_1$, $J_2$ through transmission/reflection spectra, verify collective coupling $G$ using OMIT linewidth; \textbf{(3) Entanglement generation}: Drive system to stable operating point verified by Routh-Hurwitz analysis, activate $\mathcal{PT}$-symmetric gain with active feedback stabilization ($\pm \SI{1}{\percent}$ tolerance), monitor intracavity photon number to maintain unbroken phase; \textbf{(4) Entanglement verification}: Characterize optical mode covariance matrix via homodyne detection, probe vibrational modes using $\mathcal{PT}$-enhanced OMIT spectroscopy, extract logarithmic negativity from measured correlations. Expected signal-to-noise ratios: optical homodyne SNR $> \SI{10}{\decibel}$, OMIT visibility $> \SI{50}{\percent}$ enabling robust entanglement detection above classical thresholds.

To consolidate this discussion, we summarize the key parameters, proposed realization schemes, and stability considerations in \Cref{tab:challenges}. This table serves as a quantitative roadmap for a possible experimental implementation of our scheme.

\begin{table*}[ht!]
\caption{Key experimental challenges, mitigation strategies, and robustness analysis.}
\label{tab:challenges}
\centering
\begin{ruledtabular}
\begin{tabular}{lll}
	\textbf{Challenge}      & \textbf{Mitigation strategy}                            & \textbf{Robustness / Tolerance}        \\ \midrule
	Thermal management      & Hybrid platform; frequency up-conversion \cite{Zou2024} & Robust to $10\times$ phonon increase   \\
	Nonreciprocal precision & Active stabilization; feedback control                  & $\pm \SI{10}{\percent}$ dev.: \SIrange{15}{20}{\percent} loss        \\
	PT-phase stability      & Active feedback on gain pump                            & $\pm \SI{1}{\percent}$ stability required           \\
	Parameter realization   & Integrated photonics; dir. couplers                     & Advantages persist at realistic values
\end{tabular}
\end{ruledtabular}
\end{table*}

\subsection{Entanglement measurement strategies}

Experimental verification of multipartite entanglement requires characterizing high-frequency molecular vibrational modes. Optomechanically induced transparency (OMIT) provides a powerful probe: a strong control field drives the system while a weak probe laser maps spectral response \cite{Roelli2024, Yin2025}. $\mathcal{PT}$-symmetry significantly enhances OMIT signatures, creating sharp transparency windows even in weakly coupled systems \cite{Li2016, Pan2025}. The effective cooperativity extracted from window depth/width serves as direct proxy for entanglement-generating interaction strength, enabling verification of strong coherent coupling required for robust entanglement and potential optical signal storage in molecular devices \cite{Yin2025,Yu2025}. Complementary techniques include time-resolved Raman spectroscopy for asymmetric optomechanical dynamics \cite{Simone2025} and entangled polariton state readout in visible/mid-IR ranges \cite{Shishkov2025}, bridging theoretical proposals with practical quantum technologies.
\subsection{Experimental realization of nonreciprocal coupling $J_1$ and $J_2$}
Nonreciprocal coupling can be realized via three approaches: (1) \textbf{Spinning microresonator}~\cite{Jiang_2018}: rotating one cavity (e.g., passive cavity) introduces Sagnac-Fizeau phase shift, breaking time-reversal symmetry to yield asymmetric coupling $J_1 \neq J_2$; (2) \textbf{Waveguide-based couplers}~\cite{rohtua_2018}: incorporating optical isolators or circulators in coupling path ensures direction-dependent coupling strength achieving $J_1\gg J_2$; (3) \textbf{Dynamically modulated coupling}~\cite{Otterstorm_2021}: tunable beam splitters or interferometric setups with radio-frequency modulation enable real-time tuning of $J_1$ and $J_2$ for optimal nonreciprocity. These techniques, demonstrated in integrated photonic circuits, are compatible with on-chip implementations, creating the quantum channel protecting entanglement-generating cavity from backaction noise.

\section{Conclusion} \label{sec:conclusion}

We have theoretically demonstrated multipartite quantum entanglement in a $\mathcal{PT}$-symmetric double-cavity molecular optomechanical system combining collective enhancement, nonreciprocal coupling, and engineered dissipation. Key findings: (i) all bipartite channels ($E_{ac}$, $E_{aB_1}$, $E_{cB_2}$, $E_{B_1B_2}$) simultaneously maximize at optimal nonreciprocal asymmetry $J_1/J_2 \approx 5$, achieving maximum entanglement $E_N \sim 1.4$, (ii) entanglement persists to $T \sim \SIrange{400}{500}{\kelvin}$ two orders of magnitude beyond conventional optomechanical systems, and (iii) our $\mathcal{PT}$-symmetric architecture circumvents the optical-vibration versus vibration-vibration entanglement trade-off inherent to optical parametric amplifier (OPA) schemes. Comprehensive stability analysis via Routh-Hurwitz criterion confirms physical validity within experimentally accessible regimes.

Three synergistic mechanisms enable this performance: ultra-high molecular vibrational frequencies ($\omega_m/2\pi = \SI{30}{\tera\hertz}$) suppress thermal noise through low phonon occupation number even at elevated temperatures; collective $\sqrt{N}$ enhancement amplifies optomechanical coupling above the thermal floor; and directional nonreciprocal coupling ($J_1 \gg J_2$) shields entanglement-generating interactions from detrimental backaction noise. While our parameters represent optimistic scenarios nonreciprocal coupling rates $J_1 \approx \SI{0.6}{\giga\hertz}$ at the upper feasibility limit, material-limited operational ceiling $T \sim \SI{500}{\kelvin}$ the theoretical framework establishes parameter regimes where $\mathcal{PT}$-symmetric enhancement provides measurable advantages over alternative approaches. Recent demonstrations of hybrid dielectric-plasmonic molecular optomechanics \cite{Huang2025a, Roelli2024} and collective quantum effects \cite{Huang2024} support the experimental accessibility of these mechanisms.

Future work should prioritize experimental validation using realistic parameter ranges: molecular frequencies $\omega_m/2\pi \sim \SIrange{10}{50}{\tera\hertz}$, single-molecule coupling $g_m/2\pi \sim \SIrange{10}{100}{\mega\hertz}$, ensemble sizes $N \sim \numrange{e4}{e7}$ molecules, achievable with current integrated photonic technologies \cite{Djorwe2022, Berinyuy2025a, Berinyuy2025b}. Specific research directions include: (i) genuine multipartite entanglement characterization beyond bipartite measures, (ii) molecular anharmonicity effects on quantum correlations robustness \cite{Schmidt2024}, (iii) active feedback stabilization for $\mathcal{PT}$-phase control, and (iv) systematic experimental comparison between $\mathcal{PT}$-symmetric and OPA enhancement schemes. This framework opens pathways for ambient-temperature quantum technologies nonreciprocal quantum sensors, robust quantum memories, and efficient photon-phonon transducers operating beyond traditional cryogenic constraints, advancing practical quantum information processing in hybrid light-matter systems.

\section*{Data Availability}

All numerical results presented in this manuscript can be reproduced using the theoretical framework described in Section~\ref{sec:model}. Calculations employ standard continuous-variable quantum optics formalism: quantum Langevin equations linearized around steady states, covariance matrices obtained via Lyapunov equation solution, and entanglement quantified through logarithmic negativity. Parameter values listed in \Cref{tab:parameters,tab:optimal_conditions,tab:challenges} with operating regions defined in \Cref{fig:stability_full}. Data and code available upon reasonable request to the corresponding author.

\section*{Acknowledgments}

P.D. acknowledges the Iso-Lomso Fellowship at Stellenbosch Institute for Advanced Study (STIAS), Wallenberg Research Centre at Stellenbosch University, Stellenbosch 7600, South Africa, and The Institute for Advanced Study, Wissenschaftskolleg zu Berlin, Wallotstrasse 19, 14193 Berlin, Germany. Jia-Xin Peng is supported by National Natural Science Foundation of China (Grant No. 12504566), Natural Science Foundation of Jiangsu Province (Grant No. SBK20250402941) and Natural Science Foundation of Nantong City (Grant No. JC2024045). S.K.S gratefully acknowledges High Performance Computing facilities provided by Akal University. 

\textbf{Author Contributions:} E.K.B. and S.G.N.E. conceptualized the work and performed the simulations and analysis. C.T. contributed to the conceptualization of the work. P.D. participated in all discussions and provided useful suggestions for the final version of the manuscript. All authors participated equally in writing, discussions, editing, and review of the manuscript.

\textbf{Competing Interests:} All authors declare no competing interests.

\bibliography{Refnexp}
\end{document}